\documentclass[11pt,a4paper]{article}
\pdfoutput=1
\usepackage{jheppub}
\usepackage[utf8x]{inputenc}
\usepackage{amsmath}
\usepackage{amsfonts}
\usepackage{amssymb}
\usepackage{graphicx}
\usepackage{bbm}
\usepackage{xspace}

\newcommand{\SU}{\text{SU}}
\newcommand{\U}{\text{U}}
\newcommand{\dd}{D}
\newcommand{\one}{\mathbbm{1}}
\newcommand{\Dtwo}[1]{{D_{#1}^{(N_f=2)}}}
\newcommand{\DOT}{.}
\newcommand\qone{QCD$_1$\xspace}
\newcommand\cc{\text{c.c.}}
\newcommand\mut{\mu}
\newcommand\muct{\mu_c}
\newcommand\mupmp{\pm\mu_c+\mu}
\newcommand\mupmm{\pm\mu_c-\mu}
\newcommand\mumpp{\mp\mu_c+\mu}
\newcommand\mumpm{\mp\mu_c-\mu}
\newcommand\mump{(-\mu_c+\mu)}
\newcommand\mumm{(-\mu_c-\mu)}
\newcommand\bset[2]{\underset{\makebox[0mm]{\scriptsize$#1$}}{\underbrace{#2}}}
\newcommand\ev[1]{\langle{#1}\rangle}
\newcommand\evz[1]{\langle{#1}\rangle_{{}_0}}

\DeclareMathOperator{\re}{Re}
\DeclareMathOperator{\im}{Im}
\DeclareMathOperator{\diag}{diag}
\DeclareMathOperator{\sign}{sign}
\DeclareMathOperator\arsinh{arsinh}
\DeclareMathOperator\tr{tr}

\title{Subset method for one-dimensional QCD}

\author{Jacques Bloch, Falk Bruckmann, and Tilo Wettig}
\affiliation{Institute for Theoretical Physics, University of
  Regensburg, 93040 Regensburg, Germany}

\emailAdd{jacques.bloch@ur.de}
\emailAdd{falk.bruckmann@ur.de}
\emailAdd{tilo.wettig@ur.de}

\abstract{We present a subset method which solves the sign problem for
  QCD at nonzero quark chemical potential in 0+1 dimensions. The
  subsets gather gauge configurations based on the center symmetry of
  the SU(3) group. We show that the sign problem is solved for one to
  five quark flavors and that it slowly reappears for a larger number
  of flavors. We formulate an extension of the center subsets that
  solves the sign problem for a larger number of flavors as
  well.  We also derive some new analytical results for this toy
  model.}


\begin{document}

\maketitle
\flushbottom

\allowdisplaybreaks[2]

\section{Introduction}

Numerical simulations of quantum chromodynamics (QCD) at nonzero
quark chemical potential are seriously hampered by the sign problem,
caused by the fluctuating sign of the fermion determinant. Although
this sign problem is particularly serious in the four-dimensional
theory, it is already apparent in (0+1)-dimensional QCD (\qone)
\cite{Bilic:1988rw}.  Thus \qone can be used as a toy model to study
the sign problem \cite{Ravagli:2007rw}, which is the main focus of the
present paper.  The sign problem in \qone is actually mild such that
reweighting methods can be used in simulations.  Nevertheless, it is
worthwhile to attempt to solve the sign problem exactly since such a
solution could help us to better cope with the sign problem in
higher-dimensional gauge theories.

A very general method that has been used in the past to solve or
alleviate a number of sign problems can be called ``subset method''.
In this method, configurations appearing in the ensemble are grouped
into subsets such that the sum of the weights associated with these
configurations is real and positive.  This positivity implies that
Markov chains of relevant subsets can be generated using importance
sampling methods in Monte Carlo simulations.  For example, in the
dimer algorithm \cite{Rossi:1984cv,Karsch:1988zx}, which is a
reformulation of strong-coupling QCD, dimer and baryon loops are
gathered into subsets in order to alleviate the sign problem.  Subsets
in which the configurations are related by $Z_2$ or $Z_3$ rotations
were introduced to solve the sign problem exactly in simulations of
spin models using cluster algorithms
\cite{Chandrasekharan:1999cm,Alford:2001ug} and to alleviate the sign
problem in QCD
\cite{Barbour:1988ax,Hasenfratz:1991ax,Aarts:2001dz,Bringoltz:2010iy}.
Recently, a subset method was proposed that solves the sign problem in
simulations of a random two-matrix model of QCD
\cite{Bloch:2011jx,Bloch:2012ye}.  In this method, the configurations
in a subset are related by orthogonal rotations, and it has since been
shown that the subset positivity is closely related to a projection on
the canonical determinant with zero quark number \cite{Bloch:2012bh}.
Since even in 0+1 dimensions QCD has a richer structure than just the
canonical partition function with zero quark number, this particular
subset method is not applicable to QCD.

In this work we investigate the idea to construct subsets with real
and positive weights in the context of QCD, and specifically focus on
\qone.  Our present application of the subset idea to an SU(3) gauge
theory is fundamentally different from the subset method applied to
random matrix theory. Indeed, the rotations applied in the latter are
not allowed in QCD as they would move the configurations outside the
simulated theory.  Instead, the subset construction proposed here is
based on the $Z_3$ center symmetry of the SU(3) group.  As mentioned
above, the idea of $Z_3$ averaging has been successfully applied in
other theories before.  In our case of \qone, pure $Z_3$ rotations solve
the sign problem, but only for a small number of flavors.  We then
introduce an extension of the subset construction that solves the sign
problem also for a higher number of flavors.\footnote{Note that for a
  single flavor the sign problem in \qone could also be solved exactly
  using the dimer algorithm \cite{Rossi:1984cv,Karsch:1988zx},
  although this method was not yet explicitly applied to \qone.}

Note that there is also a severe sign problem in the U(3) theory in
one dimension \cite{Ravagli:2007rw,Lombardo:2009aw,Aarts:2010gr}. However, this sign problem can
readily be solved as the subset method developed for random matrix
theory can be directly ported to this gauge theory. We will not
consider this theory further in this work as its physical content,
i.e., the lack of baryons, is clearly different from that of QCD.

The structure of this paper is as follows.  In section~\ref{sec:1dqcd}
we briefly review \qone.  In section~\ref{sec:subset_partfunc} we
present the subset construction based on pure $Z_3$ rotations for a
single flavor and show analytically and through numerical simulations
that it solves the sign problem.  In section~\ref{sec:subset_flavors}
we apply the subset method to a larger number of flavors and observe
that the sign problem reappears for six or more flavors.  We show how
the remaining sign problem can be solved by an extension of the subset
construction.  A summary is given in section~\ref{sec:summ}.  We also
compute a number of new analytical results, some of which are used to
check the numerical simulations.  Details on their derivation are
provided in several appendices.

\section{QCD in 0+1 dimensions}
\label{sec:1dqcd}

The system we analyze is an SU(3) gauge theory on a lattice with
zero spatial extent and $N_t$ sites in the temporal direction, whose
physical extent is the inverse temperature $1/T=N_t a$, with $a$ the
lattice spacing.  The one-dimensional Dirac operator for a quark of
mass $m$ at chemical potential $\mu$ reads \cite{Bilic:1988rw}
\begin{align}
aD =
\begin{pmatrix}
am & e^{a\mu} U_1/2 & 0 & \cdots & 0 & e^{-a\mu}U_{N_t}^\dagger/2 \\
-e^{-a\mu} U_1^\dagger/2 & am & e^{a\mu} U_2/2 & \cdots & 0 & 0 \\
\vdots & \vdots & \vdots & \ddots & \vdots &\vdots \\
0 & 0 & 0 & \cdots & am & e^{a\mu} U_{N_t-1}/2 \\
-e^{a\mu}U_{N_t}/2 & 0 & 0 & \cdots & -e^{-a\mu} U_{N_t-1}^\dagger/2 & am
\end{pmatrix} ,
\label{D1d}
\end{align}
where $U_1,\ldots, U_{N_t}$ are the gauge links, and the opposite sign
in front of $U_{N_t}$ and $U_{N_t}^{\dagger}$ accounts for the
antiperiodic boundary conditions of the fermions in the temporal
direction.  In the following we assume that $N_t$ is even.

Via a gauge transformation all links in the temporal direction can be
shifted into a single link, $U_1\cdots U_{N_t} \equiv P$, where $P$ is
the Polyakov loop. Due to the low dimensionality, there is no field
strength (i.e., no plaquette) and thus no gauge action. The partition
function
\begin{align}
 Z^{(N_f)}=\int dP\,{\det}^{N_f}[aD(P)]\,,
 \label{Z1dQCD}
\end{align}
where $dP$ is the Haar measure of SU(3), is thus simply a one-link
integral of the determinant of the Dirac operator for $N_f$ quark
flavors, which for simplicity we take to be degenerate, coupled to a
chemical potential $\mu$.  Expectation values of observables are
defined in the usual way,
\begin{align}
  \langle O \rangle= \frac{1}{ Z^{(N_f)}} \int dP\,
  {\det}^{N_f}[aD(P)] \, O(P)\,.
\end{align}

One can show that (for even $N_t$) the Dirac determinant can be
reduced to the determinant of a $3\times 3$ matrix \cite{Bilic:1988rw},
\begin{align}
  \det (aD)=\frac{1}{2^{3N_t}}\det (A\,\one_3+e^{\mu/T} P+e^{-\mu/T}
  P^\dagger) 
  \label{detD}
\end{align}
with 
\begin{align}
  A=2\cosh\left(\mu_c/T\right)  
\end{align}
and critical chemical potential \cite{Bilic:1988rw,Ravagli:2007rw}
\begin{align}
  \label{eq:muc}
  a\mu_c = \arsinh(am)\,.
\end{align}
The chiral limit $m=0$ correspond to $A=2$.  In the continuum limit $a
\to 0$ one has $\mu_c\to m$.  Equation~\eqref{detD} shows that
$\det(aD)$ depends on $P$ and $\mu$ only through the combination
$e^{\mu/T} P$.  This is due to the facts that (i) all gauge links can
be shifted into one, which then equals $P$, and (ii) the Dirac
determinant depends on $\mu$ only through closed temporal loops, which
give rise to a factor of $(e^{a\mu})^{N_t}=e^{\mu/T}$.  From now on we
set $a=1$.

The determinant \eqref{detD} can be decomposed into powers of $e^{\mu/T}$ as
\begin{align}
 \det D(P)
 &= \sum_{q=-3}^3 \dd_{q}(P) \, e^{q\mu/T} ,
\label{eq:decomp}
\end{align}
where the coefficients $D_q$ are the canonical
determinants.\footnote{For general gauge group SU($N_c$) the lowest
  and highest powers are $\pm N_c \mu/T$.}  From now on we omit the
irrelevant prefactor $1/2^{3N_t}$ in \eqref{detD} such that the first
and the last term on the RHS of \eqref{eq:decomp} have unit
coefficients, i.e., $\dd_{-3}=\dd_3=1$. The coefficients $\dd_{q}$ are
given in appendix~\ref{app:detD}. They depend on the configuration,
i.e., the Polyakov loop, and are generically complex. From
eq.~\eqref{detD} we see that $\det D(\mu) = \det D^\dagger(-\mu)$ such
that the canonical determinants satisfy $\dd_q^*=\dd_{-q}$. The
imaginary parts of the coefficients $D_q$ can be cancelled by pairing
each Polyakov loop with its complex conjugate (in the same spirit as
in the subsets below), as $\det D(P^*)=[\det
D(P)]^*$.\footnote{\label{foot}In \qone we could alternatively pair
  $P$ with $P^\dagger$ instead of $P^*$ since $\det D(P^\dagger)=[\det
  D(P)]^*$ in this case, see \eqref{detD}.  However, in higher
  dimensions $\det(\{U^\dagger\})\ne[\det D(\{U\})]^*$, while $\det
  D(\{U^*\})=[\det D(\{U\})]^*$ always holds.}  The sum of the
determinants for $P$ and $P^*$ is real, but without a definite
sign. The fact that these real parts can have fluctuating signs causes
the sign problem in probability-based approaches to the path
integral. At vanishing $\mu$ the Dirac operator has complex conjugate
pairs of eigenvalues such that its determinant is real and
nonnegative.

\section{\boldmath Subset method for $N_f=1$}
\label{sec:subset_partfunc}

\subsection{Subset construction and properties}
\label{sec:subsets}

We first demonstrate the subset method for the one-flavor partition
function. The aim of the subset method is to gather configurations of
the ensemble into small subsets such that the sum of their weights
contributing to the partition function is real and nonnegative. This
basic idea is identical to that proposed for random matrix theory in
refs.~\cite{Bloch:2011jx,Bloch:2012ye}, but the generation of the
subsets will be fundamentally different, mainly because of the
stringent constraint that the configurations be elements of the
SU(3) gauge group.

In the proposed subset method, starting from a seed configuration $P$
a subset $\Omega_P$ is formed containing three SU(3)
elements:\footnote{For gauge group $\SU(N_c)$ the subset method is
  generalized in a straightforward way with $N_c$ configurations in
  each subset.} the seed configuration itself and the SU(3) elements
generated by rotating the seed by the two nontrivial center elements
of SU(3),
\begin{align}
  \Omega_P = \{ P, e^{2\pi i/3} P, e^{4\pi i/3} P \} \,.
  \label{eq:Pset}
\end{align}
For any $P \in \SU(3)$, the rotated configurations are again elements
of $\SU(3)$ and thus part of the ensemble. As the subsets are
invariant under $Z_3$ rotations, the set of all subsets forms a
three-fold covering of the original SU(3) ensemble. Therefore, we
define the subset weights as
\begin{align}
 \sigma(\Omega_P)=\frac{1}{3}\sum_{k=0}^{2}\det D(P_k)
\label{eq:subset_def}
\end{align}
with $P_k=e^{2\pi i k/3}P$.  Note that any of the three configurations
can be used as a seed of the subset. As the Haar measure $dP$ is
invariant under the center rotations, the \qone partition function can
be rewritten as an integral over the subsets,
\begin{align}
 Z^{(1)}=\int dP \,\sigma(\Omega_P) \,.
\label{eq:Z_rewritten}
\end{align}
To compute observables one has to take into account that the three
configurations in a subset can have different values of the observable,
such that
\begin{align}
\langle O \rangle
= \frac{1}{Z^{(1)}} \int dP \,\sigma(\Omega_P) \, \langle O \rangle_{\Omega_P} 
\label{<O>}
\end{align}
with subset measurements
\begin{align}
  \langle O \rangle_{\Omega_P}= \frac{1}{3\sigma(\Omega_P)}
  \sum_{k=0}^{2}\det D(P_k) \, O(P_k) \,, 
  \label{<O>sigma}
\end{align}
where $P_k \in \Omega_P$ and $\langle 1 \rangle_\Omega=1$. The measure
$dP\,\sigma(\Omega_P)$ in eq.~\eqref{<O>} indicates that subsets of
configurations, rather than individual configurations, will be
generated in the numerical simulations, such that observables will be
approximated as sample means of $N_\text{MC}$ subset measurements,
\begin{align}
\overline O = \frac{1}{N_\text{MC}} \sum_{n=1}^{N_\text{MC}} \langle O \rangle_{\Omega_{n}} \,.
\label{Obar}
\end{align}

The subset weights $\sigma$ are the main ingredients of the method, and
their properties will now be analyzed further. From eq.~\eqref{detD}
we observe that multiplying the Polyakov loop by a phase factor
$e^{i\theta}$ can be reinterpreted as adding an imaginary part
$i\theta$ to the chemical potential $\mu/T$ \cite{Bloch:2012bh},
\begin{align}
 \det D(e^{i\theta}P)\big|_{\mu/T}=\det D(P)\big|_{\mu/T+i\theta} \,.
 \label{Z3tomu}
\end{align}
While this relation holds for any angle $\theta$, we will only use
$\theta=0,\,2\pi /3,\,4\pi/3$ for the rotations of $P$ in order to
remain in the gauge group.  By applying eq.~\eqref{Z3tomu} to the
subset weight \eqref{eq:subset_def} we find that the sum of
determinants in a subset effectively projects the determinant
\eqref{eq:decomp} onto its components with zero triality ($q\bmod
3=0$),\footnote{Here we see the difference to the random-matrix model
  \cite{Bloch:2011jx,Bloch:2012ye,Bloch:2012bh} and gauge theories
  with gauge group U($N$), see, e.g., \cite{Langfeld:2011rh}.  In
  those cases, the configuration space allows for rotations by
  arbitrary angles $\theta$ such that the subsets project onto the
  sector with vanishing quark number.  As a result, the partition
  function does not depend on $\mu$ at all.}
\begin{align}
 \sigma(\Omega_P)=\frac{1}{3}\,\sum_{k=0}^2\,\sum_{q=-3}^3 
 \dd_q(P) \, e^{q(\mu/T+2\pi i k/3)}
 =\sum_{b=-1}^1 \dd_{3b}(P)\,e^{3b\mu/T} \,,
\label{eq:subset_expansion}
\end{align}
where in the last step we changed the summation index from $q$
(``quark number'') to $b$ (``baryon number'').  The last equality
follows from the well-known formula for the sum of the $q$-th powers
of the $N$-th roots of unity, $\sum_{k=0}^{N-1}\exp(2\pi i q k/N)=N
\delta_{q\bmod N,0}$.  As the same expansion
\eqref{eq:subset_expansion} can be derived using any of the three
subset elements as the seed for the subset, it follows that the
canonical determinants with triality zero are identical for the three
subset elements.  In section~\ref{sec:subset_flavors} we will show
that this projection cures, or at least attenuates, the sign problem
depending on the number of flavors being considered.

As the partition function can be written as an integral over the
subsets, we obtain from \eqref{eq:subset_expansion} the fugacity
expansion
\begin{align}
 Z^{(1)}(\mu)=\sum_{b=-1}^1 Z^{(1)}_{3b}\,e^{3b\mu/T}
 \label{eq:fugacity}
\end{align}
with canonical partition functions
\begin{align}
 Z^{(1)}_q = \int dP\,\dd_q(P) \,.
\end{align}
The absence of the other triality contributions ($q\bmod3\ne0$) is
a direct consequence of the center symmetry of SU(3). Although the
determinants of the individual configurations contribute to all
triality sectors, the subset approach, by making use of the center
symmetry, automatically projects onto the only triality sector that
contributes to the overall partition function.

Let us rewrite eq.~\eqref{eq:subset_expansion} as
\begin{align}
 \sigma(\Omega_P)=\dd_{0}(P)+2\cosh(3\mu/T) \,.
 \label{sigma}
\end{align}
The second term is obviously positive, and from \eqref{D0} the
constant term is
\begin{align}
  \dd_0(P)=A^3+A(|\tr P|^2-3)
\end{align}
with $A=2\cosh(\mu_c/T) \geq 2$ and $|\tr P|\in[0,3]$ such that
$\dd_0$ is positive, too. Therefore, the subset weight
$\sigma(\Omega_P)$ is real and positive for any $m$, $\mu$, and $P$
and can be used to generate subsets with importance sampling in Markov
chain Monte Carlo methods.

\subsection{Partition function and observables}

Once we have formulated the partition function in terms of subsets, it
can be computed analytically by performing a group integration of
eq.~\eqref{sigma}, which yields
\begin{align}
  \label{eq:Zmu}
  Z^{(1)}=A^3-2A+2\cosh(3\mu/T) \,,
\end{align}
where we used $\int dP\, |\tr P|^2=1$ as shown in
appendix~\ref{app:traces}. Using the definition of $A$ this can be
rewritten as
\begin{align}
  Z^{(1)} = \frac{\sinh(4\mu_c/T)}{\sinh(\mu_c/T)} + 2\cosh(3\mu/T) \,,
 \label{Zm}
\end{align}
which agrees with the literature \cite{Bilic:1988rw,Ravagli:2007rw}.
From the partition function \eqref{eq:Zmu} we can easily compute the
chiral condensate as its mass derivative,
\begin{align}
  \Sigma = T \frac{\partial\log Z^{(1)}}{\partial m}
  =\frac{1}{Z^{(1)}}\frac{4\sinh(\mu_c/T)}{\cosh(a\mu_c)}
  \big[5+6\sinh^2(\mu_c/T)\big]\,,
  \label{ccth}
\end{align}
where we have used \eqref{eq:muc} and temporarily restored the lattice
spacing $a$.  Note that the discretization-dependent factor
$\cosh(a\mu_c)$ goes to one in the continuum limit.  From \eqref{ccth}
we see that the $\mu$-dependence of the chiral condensate is given by
the inverse partition function, so asymptotically it behaves like
$e^{-3|\mu|/T}$.

The quark number density is computed by taking the derivative of the
partition function with respect to the chemical potential,
\begin{align}
  n = T\frac{\partial\log Z^{(1)}}{\partial\mu} =
  \frac{6\sinh(3\mu/T)}{Z^{(1)}} \,.
  \label{qnth}
\end{align}
For large $\mu$ we find the expected saturation
\begin{align}
  \lim_{\mu\to\pm\infty} n = \frac{6\sinh(3\mu/T)}{2\cosh(3\mu/T)} \to \pm 3 \,.
\end{align}

\subsection{Polyakov loop}
\label{sec:Polyakov}

Another observable of interest, which, however, cannot be computed
directly as a derivative of the partition function, is the trace of
the Polyakov loop.  Below, we show that its expectation value,
\begin{align}
 \langle \tr P\rangle=\frac{1}{Z^{(1)}} \int dP\, \det D(P) \, \tr P 
 = \frac{1}{Z^{(1)}} \int dP\, \sigma(\Omega_P) \, \langle\tr P\rangle_{\Omega_P}  
\end{align}
can elegantly be computed using the subset construction.  To do so,
one needs to evaluate the subset measurement \eqref{<O>sigma},
\begin{align}
 \langle \tr P\rangle_{\Omega_P}=\frac{1}{3\sigma}\sum_{k=0}^2 \det D(P_k) \, \tr P_k \,.
\end{align}
Under the center rotations the Polyakov loop traces transform as $\tr
P_k = e^{2\pi i k/3} \tr P$, and we compute along the lines of
\eqref{eq:subset_expansion}
\begin{align}
\langle \tr P\rangle_{\Omega_P}
 &= \frac{1}{3\sigma}\,\sum_{k=0}^2\,\sum_{q=-3}^3 
 \dd_q(P)\,e^{q(\mu/T+2\pi i k/3)}\times e^{2\pi i k/3} \tr P \notag\\
&=\frac{1}{\sigma} \left[\dd_{-1}(P)\,e^{-\mu/T}+\dd_{2}(P)\,e^{2\mu/T}\right] \tr P \,.
\end{align}
In contrast to the case of the subset weights, a different triality
sector ($q\bmod3=-1$) contributes to the subset measurement of the
Polyakov loop.  Substituting eqs.~\eqref{D1} and \eqref{D2} into
eq.~\eqref{<O>} and integrating over $P$ using
appendix~\ref{app:traces} we obtain
\begin{align}
 \langle \tr P\rangle =
 \frac{(A^2-1)\,e^{-\mu/T}+A\,e^{2\mu/T}}{A^3-2A+2\cosh(3\mu/T)} \,.
\label{eq:expv_p}
\end{align}
Similarly, the expectation value of the anti-Polyakov loop $P^\dagger$
will only contain terms with $q\bmod3=1$, and we have
\begin{align}
 \langle \tr P^\dagger\rangle
 =\frac{A\,e^{-2\mu/T}+(A^2-1)\,e^{\mu/T}}{A^3-2A+2\cosh(3\mu/T)}
 =\langle \tr P\rangle\big|_{\mu\to-\mu}\,.
\label{eq:expv_pdagger}
\end{align}
For very large chemical potential, the Polyakov loop decays
exponentially as it is suppressed by the partition function in the
denominator,
\begin{align}
 \langle \tr P\rangle\sim
 \begin{cases}e^{2\mu/T}\,, & \mu\to-\infty\,,\\
   e^{-\mu/T}\,, & \mu\to\infty\,,
   \end{cases}
   \qquad
 \langle \tr P^\dagger\rangle\sim
 \begin{cases}e^{\mu/T}\,, & \mu\to-\infty\,,\\
   e^{-2\mu/T}\,, & \mu\to\infty\,.
   \end{cases}
\label{eq:sympt_p}
\end{align}
Note that $\langle \tr P^\dagger\rangle\neq\langle \tr P\rangle$ since
the weight with which the average is performed is complex, i.e., the
real part of the expectation value of $\tr P$ is obtained by
integrating over $\re(\det D)\re(\tr P) - \im(\det D)\im(\tr P)$. For
$\tr P^\dagger$ the sign of the second term will be reversed, such
that different contributions will arise.  Note further that the
average Polyakov loop is real.  This is because $P^*$, which is also
part of the SU(3) ensemble, gives a contribution of $(\det D\tr P)^*$
that cancels the imaginary part in the average.

Since a positive quark chemical potential favors quarks over
antiquarks, their free energies should differ at nonzero chemical
potential. This is achieved by the aforementioned asymmetry between
the expectation values of $\tr P$ and $\tr P^\dagger$
\cite{deForcrand:2010ys}.  If we invert the sign of the chemical
potential the roles of quarks and antiquarks are merely interchanged,
and we obtain $\langle \tr P^\dagger\rangle_\mu = \langle \tr
P\rangle_{-\mu}$.

The quantities $e^{\mu/T}\langle\tr P\rangle$ and
$e^{-\mu/T}\langle\tr P^\dagger\rangle$ only contain triality zero
sectors, and from \eqref{qnth} we see that their difference is related
to the quark number as
\begin{align}
  \label{eq:nP1}
  n=\frac3A\langle e^{\mu/T}\tr P-e^{-\mu/T}\tr P^\dagger\rangle \,.
\end{align}
In appendix~\ref{app:nP} we derive relations between the quark number
density and the Polyakov loop for an arbitrary number of flavors. 

\subsection{Phase diagram}

We briefly discuss the phase diagram of \qone for one flavor, which
can be derived using the analytical formulas for chiral condensate,
number density, and Polyakov loop.

\begin{figure}
\centering
\begin{tabular}{c@{\hspace*{10mm}}c}
  \includegraphics{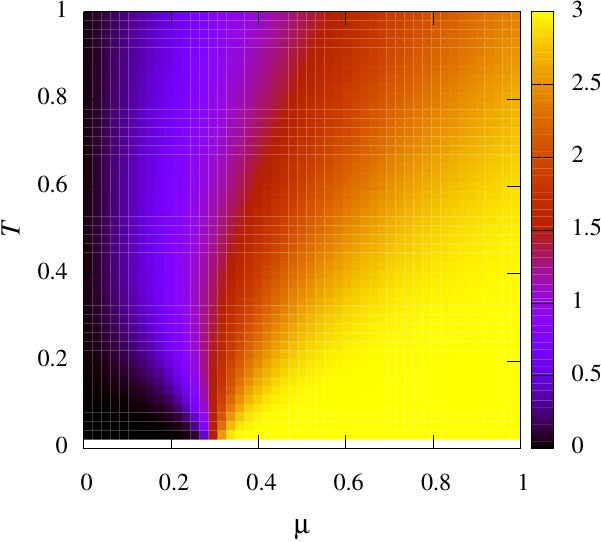}&
  \includegraphics{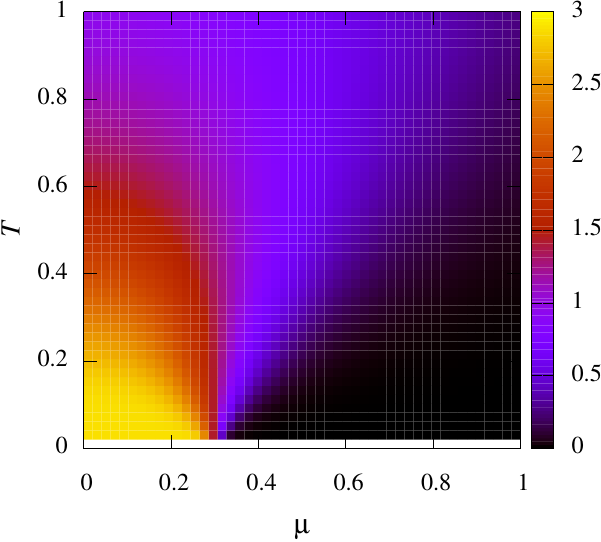}\\
  \includegraphics{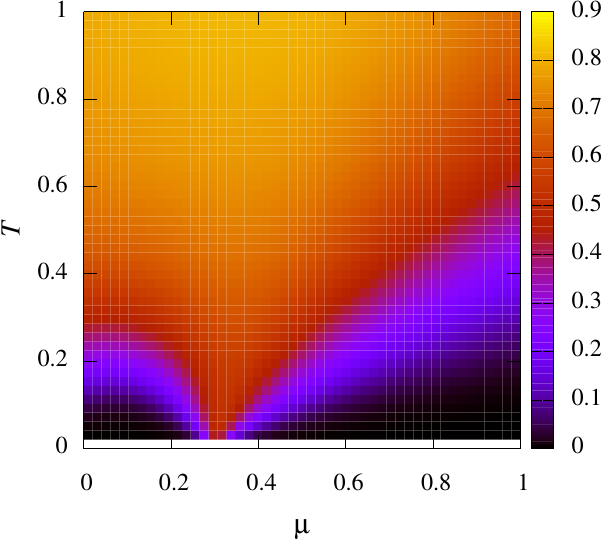}&
  \includegraphics{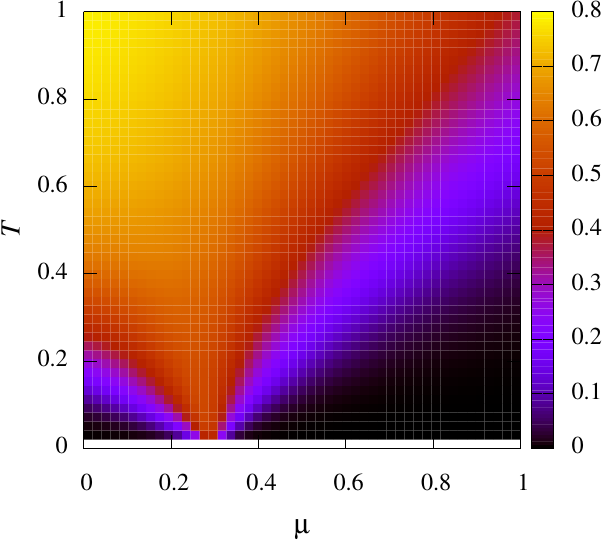}
\end{tabular}
\caption{$(\mu,T)$-diagram for the number density (top left), chiral
  condensate (top right), Polyakov loop (bottom left), and
  anti-Polyakov loop (bottom right) for $m=0.3$.}
\label{Fig:phdiag}
\end{figure}
In figure~\ref{Fig:phdiag} we show the quark number density, the
chiral condensate, and the trace of the Polyakov and anti-Polyakov
loop versus chemical potential and temperature for $m=0.3$. A true
phase transition only occurs at $T=0$ and $\mu=\mu_c$, where the
partition function (for $\mu>0$) is given by
\begin{align}
\lim_{T\to 0} Z^{(1)} = \lim_{T\to 0} ( e^{3\mu/T} + e^{3\mu_c/T}) \,.
\end{align}
From this expression we see that for $T\to0$ and $\mu<\mu_c$ the
partition function and all thermodynamic observables are independent
of $\mu$ (a fact that has been termed the Silver Blaze property
\cite{Cohen:2003kd}), while for $T\to0$ and $\mu>\mu_c$ the partition
function and all thermodynamic observables are independent of $m$
(which can be viewed as an analog of the Silver Blaze property).

\begin{figure}
\centering
\begin{tabular}{c@{\hspace*{10mm}}c}
  \includegraphics{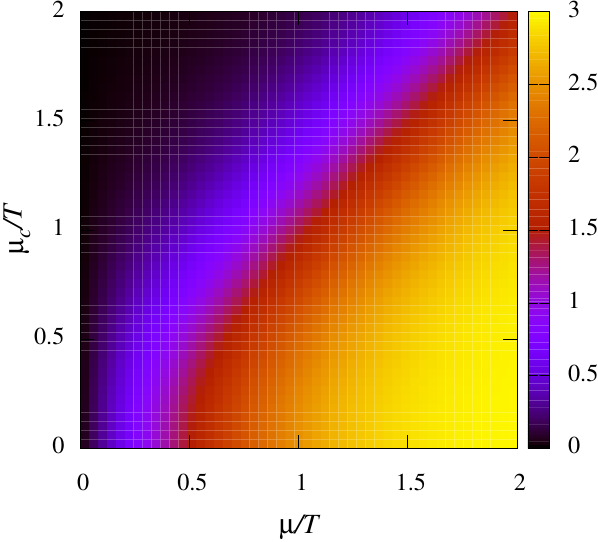}&
  \includegraphics{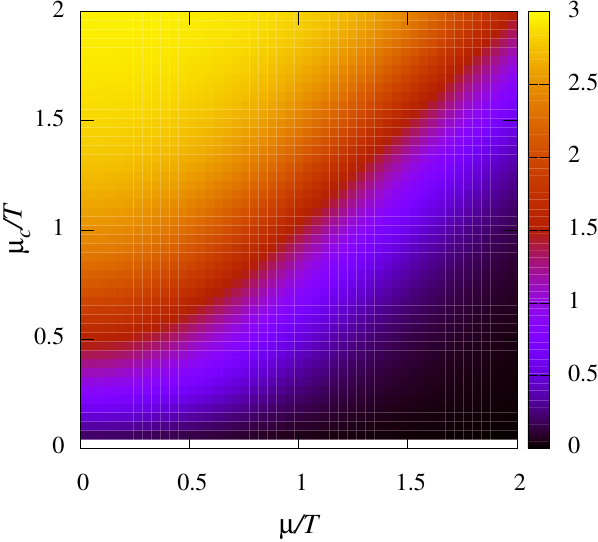}\\
  \includegraphics{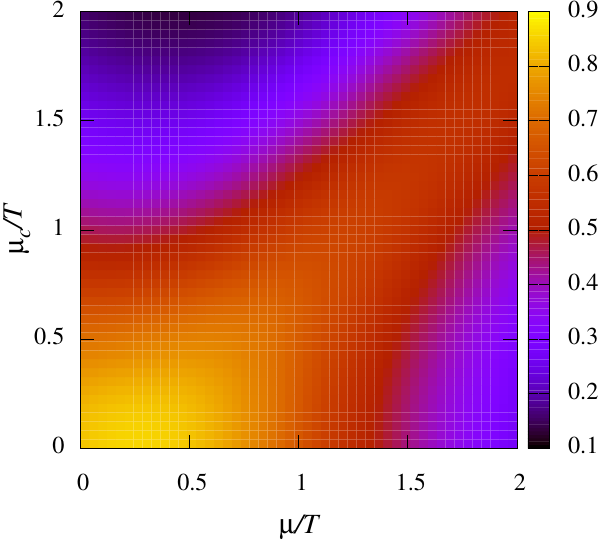}&
  \includegraphics{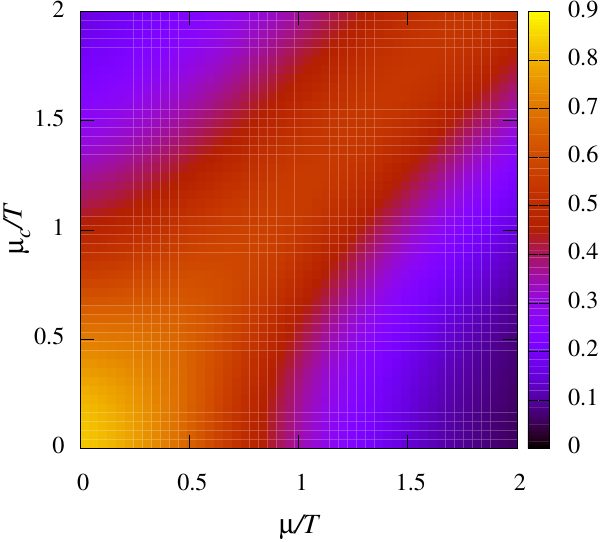}  
\end{tabular}
\caption{$(\mu/T,\mu_c/T)$-diagram for the number density (top left),
  modified chiral condensate $T\partial\log
  Z^{(1)}/\partial\mu_c=\Sigma\cosh(a\mu_c)$ (top right), Polyakov loop
  (bottom left), and anti-Polyakov loop (bottom right).}
\label{Fig:phdiag_mu_vs_muc}
\end{figure}

Note that the partition function does not depend separately on the
three variables $m$, $\mu$, and $T$ but only on the two ratios $\mu/T$
and $\mu_c/T$, where $\mu_c$ is related to $m$ via \eqref{eq:muc}.
Therefore it is interesting to look at the observables as a function
of these two scaled variables, see figure~\ref{Fig:phdiag_mu_vs_muc}.
Here we defined the modified chiral condensate $T\partial\log
Z^{(1)}/\partial\mu_c=\Sigma\cosh(a\mu_c)=\Sigma\cosh(\frac1{N_t}\,\mu_c/T)$
to eliminate an explicit dependence on $N_t$.

\subsection{Simulation results}

Although most observables in \qone can be computed by taking
derivatives of the partition function or performing the integrals over
the gauge group explicitly, the main aim of this work is to construct
a numerical method that makes it possible to perform Monte Carlo
simulations of this theory. We therefore implemented the subset method
to verify that it reliably reproduces the analytical predictions. As
the subset weights are real and positive we can generate Markov chains
of relevant subsets using the Metropolis algorithm, where the full
SU(3) links were generated according to the Haar measure using the
Bronzan algorithm \cite{Bronzan:1988wa}. We typically generated Markov
chains with 100,000 subsets. As most results only depend on $\mu/T$
and $\mu_c/T$ (except for a prefactor in the chiral condensate), the
simulations were performed using the minimally allowed time extent
$N_t=2$.\footnote{We also performed simulations with $N_t>2$ to verify
the numerical results.}

By taking the mass and chemical potential derivatives of the partition
function \eqref{Z1dQCD} we observe that the chiral condensate and
quark number density can be computed as ensemble expectation values of
the observables
\begin{align}
O_\Sigma = \frac{1}{N_t}\tr \left[D^{-1} \frac{\partial D}{\partial m}\right] = \frac{1}{N_t}\tr \left[D^{-1}\right],\qquad
O_n = \frac{1}{N_t}\tr\left[D^{-1} \frac{\partial D}{\partial\mu}\right] . 
\end{align}
To compute the expectation values with the subset method we apply
formula \eqref{Obar} with subset measurements \eqref{<O>sigma}.

\begin{figure}
\centerline{\includegraphics{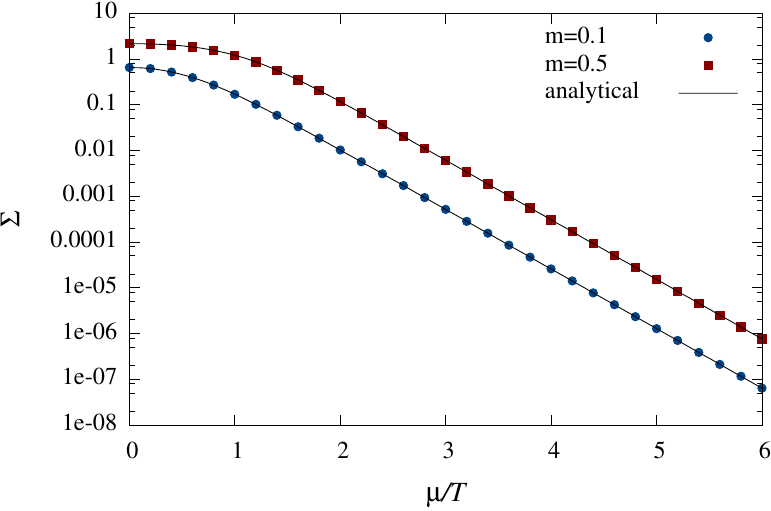}}
\caption{Chiral condensate as a function of $\mu/T$ for $N_t=2$ and
  $N_f=1$ with $m=0.1$ and $0.5$. The solid lines are the analytical
  result \eqref{ccth} from \cite{Bilic:1988rw}. The error bars are
  smaller than the symbols.\vspace*{3mm}}
\label{Fig:cc}
\end{figure}

We numerically computed the chiral condensate for different values of
$m$ using the subset method and found very good agreement with the
analytical prediction of eq.~\eqref{ccth} over several orders of
magnitude, as can be seen in figure~\ref{Fig:cc}.

\begin{figure}
\centerline{\includegraphics{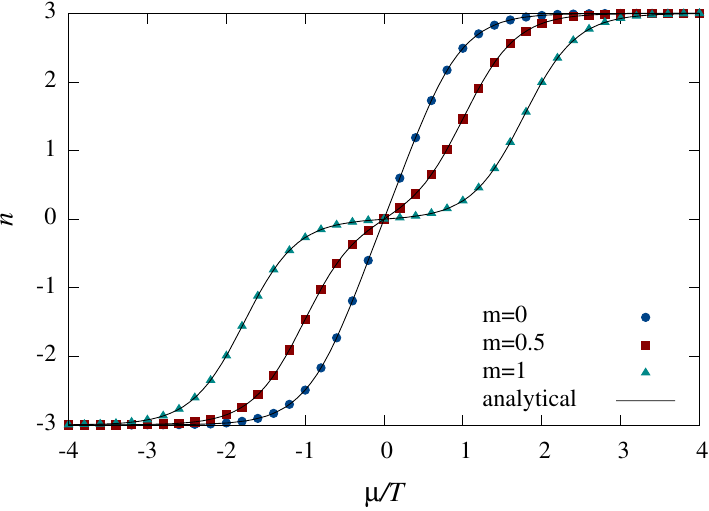}}
\caption{Quark number density as a function of $\mu/T$ for $N_f=1$
  with $m=(0,0.5,1)$. The solid lines represent the analytical
  result \eqref{qnth}.\vspace*{3mm}}
\label{Fig:qn}
\end{figure}

We also used our Monte Carlo simulations to compute the quark number
density and compared the results with the analytical prediction
\eqref{qnth} for the massless and massive cases in
figure~\ref{Fig:qn}.  As can be shown from the theoretical
formula, the linear rise at $\mu=0$ for $m=0$ is given by $n \sim 3
\mu/T$.  In the massive case, the number density still varies linearly
with $\mu$ around $\mu=0$, but the rise is much slower due to the
large denominator in \eqref{qnth}.

\begin{figure}
\centerline{\includegraphics{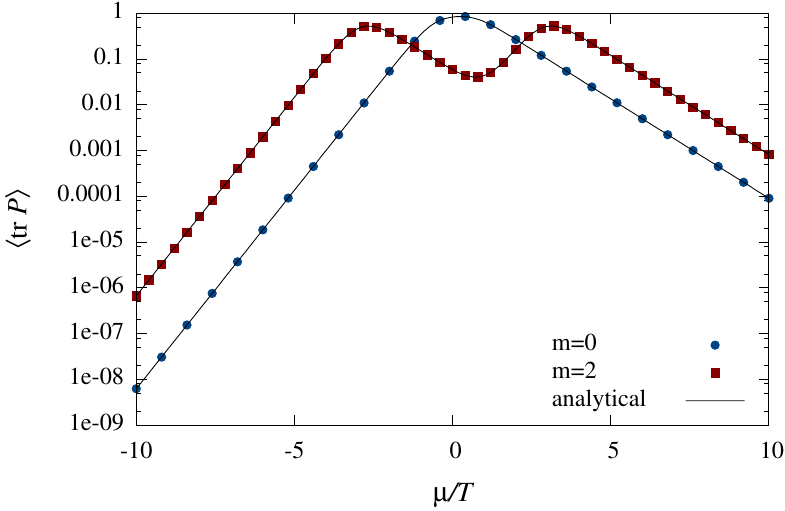}}
\caption{Average trace of the Polyakov loop $\langle\tr P \rangle$ as
  a function of $\mu/T$ for $N_f=1$ with $m=0$ and $m=2$. The solid
  lines correspond to the analytical formula \eqref{eq:expv_p}.}
\label{Fig:Ploop}
\end{figure}

Finally, we measured the average trace of the Polyakov loop and
compared the results with the prediction of eq.~\eqref{eq:expv_p} in
figure~\ref{Fig:Ploop}.  We indeed observe the $\mu \leftrightarrow
-\mu$ asymmetry (equivalent to the $\tr P \leftrightarrow \tr
P^\dagger$ asymmetry) mentioned in section~\ref{sec:Polyakov}. This is
clearly illustrated in the figure by the different exponential decays
for large positive and negative $\mu$ as described by
eq.~\eqref{eq:sympt_p}.  Measurements of the anti-Polyakov loop would
merely correspond to a $\mu \to -\mu$ exchange.

\section{\boldmath $N_f$ larger than one}
\label{sec:subset_flavors}

We now analyze the subset method for larger $N_f$. As we will see
below, the $Z_3$ subset method, introduced in section~\ref{sec:subsets}
for $N_f=1$, completely removes the sign problem for $N_f \leq 5$. The
sign problem then reappears for $N_f\geq 6$, and we discuss how it can
be solved in this case. We start the discussion with $N_f=2$.

\subsection[$N_f=2$]{\boldmath $N_f=2$}

For two flavors the fermionic weight $\det^2D$ can be decomposed
(up to an irrelevant normalization factor) into 
\begin{align}
{\det}^2D(P) = \sum_{q=-6}^6 \Dtwo{q}e^{q\mu/T} \,, 
\end{align}
where the first and the last coefficient $\Dtwo{\pm6}$ are unity
again. The $Z_3$ subsets are defined in a similar way as for $N_f=1$,
\begin{align}
  \label{eq:subscc}
  \Omega_P = \{ P, e^{2\pi i/3} P, e^{4\pi i/3} P,
  P^*, e^{2\pi i/3} P^*, e^{4\pi i/3} P^* \} \,,
\end{align}
where we now also included the complex conjugate links $P_k^*$ in
$\Omega_P$.  Their determinants satisfy $\det D(P_k^*)=[\det
D(P_k)]^*$, and thus the subset weight
\begin{align}
  \sigma_{N_f=2}(\Omega_P)=\frac{1}{6}\sum_{k=0}^{2}{\det}^2 D(P_k)
  +\cc
  \label{eq:subset_def_more}
\end{align}
is guaranteed to be real.\footnote{For $N_f=1$ (and $N_f=1$ only) the
  inclusion of the complex conjugate links $P_k^*$ was not necessary
  since in this case the subset weight was real (and positive) to
  start with.  See also footnote~\ref{foot} for the reason why we
  added $P_k^*$ and not $P_k^\dagger$.}  As before, only triality zero
contributions survive in this sum,
\begin{align}
  \sigma_{N_f=2}(\Omega_P)&=\Dtwo{0}
  +2\re\Dtwo{3}\cosh(3\mu/T)+2\cosh(6\mu/T)\,.
\label{sigmaNf=2}
\end{align}
The canonical determinants are computed by squaring
eq.~\eqref{appdetD} and substituting \eqref{D0}--\eqref{D3},
\begin{align}
  \Dtwo{0} &= D_0^2 + 2(1+|D_1|^2+|D_2|^2)\notag\\
  &= (A^6-6A^4+9A^2+2) + 4(A^4-3A^2+2)|\tr P|^2 + (A^2+2)|\tr P|^4\notag\\
  &\quad +2(A^2-2)[(\tr P)^3+(\tr P^\dagger)^3]\,,\\
  \Dtwo{3} &= 2 (D_0 + D_1 D_2) 
  = 2A\big[(A^2-3) + (A^2-1)|\tr P|^2 + (\tr P^\dagger)^3\big] \,.
\end{align}
An analysis of these expressions shows that the subset weight
$\sigma_{N_f=2}$ is positive for all Polyakov loops and all values of
chemical potential and mass, such that it can be used to generate
$N_f=2$ subsets with importance sampling.

Once we know the subset weight, the partition function readily follows by
integrating over the gauge configuration,
\begin{align}
 Z^{(2)} = (A^6-2A^4+3A^2+6) + 4A\,(2A^2-3)\cosh(3\mu/T) + 2\cosh(6\mu/T)\,,
 \label{Z2}
\end{align}
where we used the trace formulas of appendix~\ref{app:traces}.  

\begin{figure}
\centering
\includegraphics{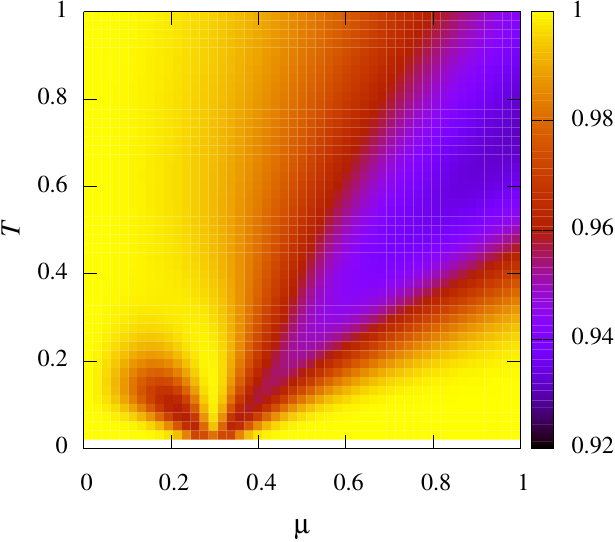}\hspace*{10mm}
\includegraphics{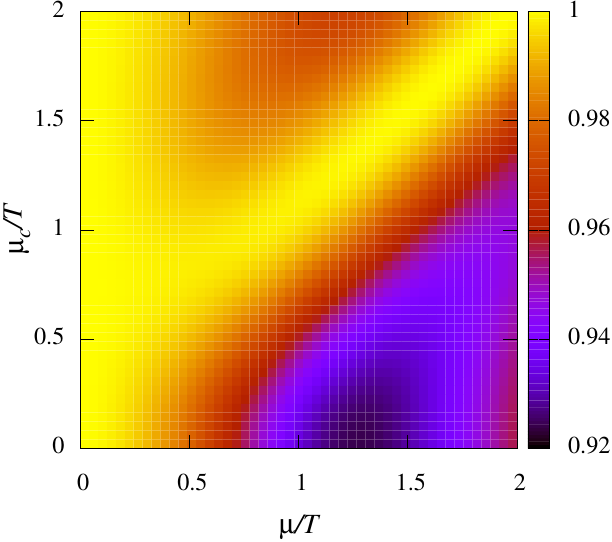}
\caption{Average phase of the fermion determinant in the
  phase-quenched ensemble for $N_f=2$, eq.~\eqref{phasepq}, as a
  function of chemical potential and temperature for $m=0.3$ (left)
  and as a function of the scaled variables $\mu/T$ and $\mu_c/T$
  (right).}
\label{Fig:theta_vs_muT}
\end{figure}

Even though the subset weights are free of the sign problem, the
determinants in the original ensemble have a fluctuating sign. This
can be well illustrated by computing the average phase of the fermion
determinant in the phase-quenched ensemble, which is also the
reweighting factor for phase-quenched reweighting (see
section~\ref{sec:largerNf}). For $N_f=2$ this average phase can be
computed analytically for arbitrary $\mu$ and $m$ and is given by
\begin{align}
  \langle e^{2i\theta} \rangle_\text{pq} 
  = \frac{Z^{(2)}}{Z^{(11^*)}} \,,
\label{phasepq}
\end{align}
where $Z^{(2)}$ is given in \eqref{Z2} and $Z^{(11^*)}=\int dP \,|\det
D|^2$ is the phase-quenched partition function, for which we find
through group integration (using either \eqref{eq:decomp} in
combination with appendix~\ref{app:detD} and \ref{app:traces}, or
\eqref{detD} in combination with the eigenvalue representation of the
Polyakov loop in appendix~\ref{app:eigrep})
\begin{align}
  Z^{(11^*)} 
  &= (A^6-4A^4+5A^2+2) + (4A^3-4A)\cosh(\mu/T) + (2A^4-4A^2+4) \cosh(2\mu/T)
  \notag\\
  &\quad + (4A^3-8A)\cosh(3\mu/T) + 2A^2\cosh(4\mu/T) + 2\cosh(6\mu/T) \,.
\end{align}
The average phase is shown in figure \ref{Fig:theta_vs_muT}.

\enlargethispage{2\baselineskip}

\subsection[Larger $N_f$]{\boldmath Larger $N_f$}
\label{sec:largerNf}

\begin{figure}\centering
\includegraphics[width=0.6\textwidth]{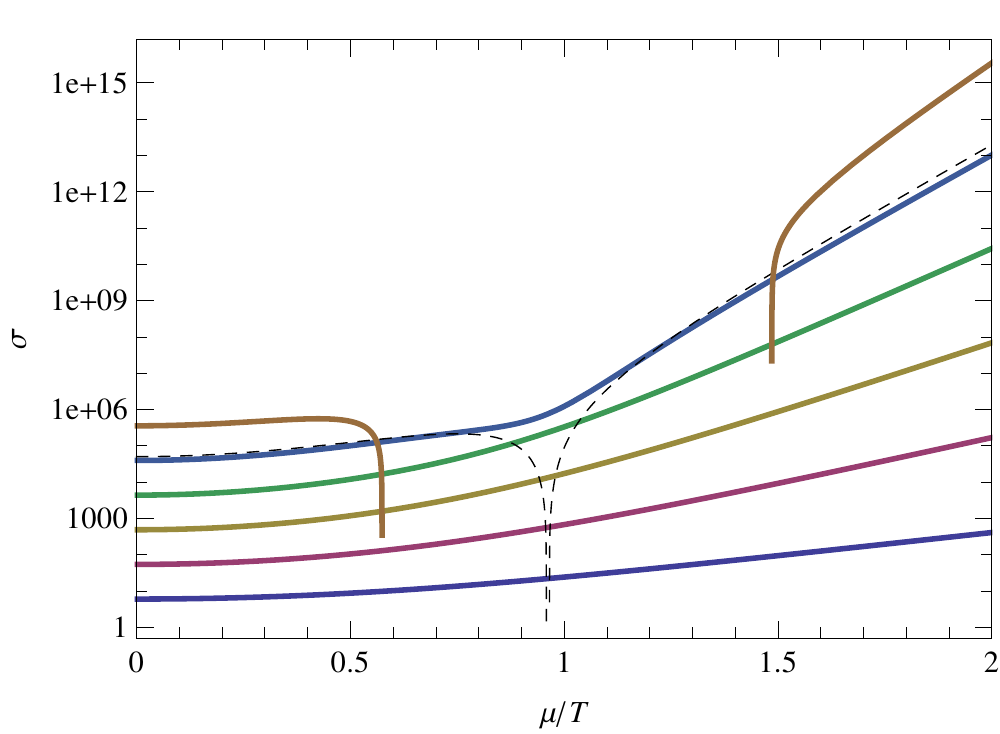}
\caption{Subset weight in the massless case for $\Omega_P$ with
  $P=\diag(1,-1,-1)$ as a function of $\mu/T$ for
  $N_f=1,2,\ldots,6$ (solid curves upwards) and a fictitious flavor
  number $N_f=5.1102$ (dashed). As the weights are plotted on a
  logarithmic scale, the gap from $\mu/T=0.57\text{ to }1.49$ for
  $N_f=6$ corresponds to negative weights.}
\label{fig_neg_first}
\end{figure}

For a larger number $N_f$ of degenerate flavors the fermionic
determinant is given by
\begin{align}
{\det}^{N_f} D(P) = \sum_{q=-3N_f}^{3N_f} D^{(N_f)}_q(P) \, e^{q\mu/T} \,,
\end{align}
where the $D^{(N_f)}_q(P)$ are the canonical determinants for $N_f$
flavors.  We construct subsets $\Omega_P$ in exactly the same way as
in eq.~\eqref{eq:subscc}. In analogy to eq.~\eqref{eq:subset_def_more}
the subset weights are
\begin{align}
  \sigma(\Omega_P) = \frac16\sum_{k=0}^2 {\det}^{N_f} D(P_k) + \cc\,,
  \label{sigmaNf}
\end{align}
and the subset measurements are given by
\begin{align}
  \label{eq:meas}
  \langle O \rangle_{\Omega_P} = \frac{1}{6\sigma(\Omega_P)}
  \sum_{k=0}^{2} \left[ {\det}^{N_f} D(P_k)\,O(P_k)
    +{\det}^{N_f} D(P_k^*)\,O(P_k^*) \right].
\end{align}
After adding the determinants of the six configurations in the
subset, we effectively project the determinants on the
triality zero sector and obtain
\begin{align}
  \sigma(\Omega_P) = D_0^{(N_f)}(P) + \sum_{b=1}^{N_f-1}
  2\re D^{(N_f)}_{3b}(P) \cosh(3b\mu/T) + 2\cosh(3N_f\mu/T)\,.
\end{align}
As in the case of $N_f=2$, the subset weight was made real by adding
the complex conjugate links to the subsets.  However, there is no
general argument for the positivity of these real subset weights for
arbitrary number of flavors. In fact, we found that the subset weights
are only strictly positive for small enough $N_f$. As we increase
$N_f$ the subset weights start to become negative.  This first happens
for $N_f\approx5.11$ at $\mu\approx0.96$ and for the subset containing
the Polyakov loop with eigenvalues $(1,-1,-1)$.  In
figure~\ref{fig_neg_first} we show the value of the subset weight for
this specific subset as a function of the chemical potential for
different numbers of flavors in the massless case.\footnote{For larger
  masses the threshold number of flavors will be higher.}  As $N_f$ is
increased further, the regions in configuration space and chemical
potential where the weights are negative slowly grow, see also
figure~\ref{fig:sigma_Nf=24} below.

Since for larger $N_f$ the subset weights are not positive on the
complete configuration space the subsets cannot be used in importance
sampling and the subset method as such no longer works. Nonetheless,
the subsets can still be useful in providing a good auxiliary system
to simulate \qone with $N_f$ flavors using reweighting methods. In
this case one generates relevant subsets according to the absolute
value of the subset weights and absorbs the sign in the
observable. The expectation value of an observable in the target
ensemble is then given by the ratio of its signed expectation value
and the average subset sign, both measured in the auxiliary ensemble,
i.e.,
\begin{align}
  \label{eq:sqrsm}
  \ev O=\frac{\ev{(\sign\sigma)\ev{O}_{\Omega_P}}_{|\sigma|}}
  {\ev{\sign\sigma}_{|\sigma|}}\,.
\end{align}
This is the so-called sign-quenched reweighting
scheme applied to the subset method.

Similar auxiliary systems can be considered for the original
formulation of the partition function in terms of the SU(3) links,
i.e., without subsets. Below, we compare the reweighting factors of
the different approaches to investigate if the subset formulation
brings an improvement to the sign problem. When comparing reweighting
schemes it is customary to compare their reweighting factors, as these
are good indicators of the severity of the sign problem and enter the
computation of all expectation values.

In the subset formulation, the average reweighting factor in the
sign-quenched reweighting scheme is the average subset sign in the
sign-quenched ensemble,
\begin{align}
 R^\sigma_{\text{sq}}
  =\langle \sign\sigma\rangle_{|\sigma|}\,.
\end{align}
As discussed above, a detailed investigation showed that all the
subset weights are positive and the reweighting factor is exactly
unity for $N_f\leq 5$. Clearly, no reweighting is necessary when using
the subset method in this case. From $N_f=6$ on, the average
reweighting factor can become smaller than unity for some range of
$\mu$.

For the original link formulation of the partition function, several
kinds of reweighting are possible, e.g., phase-quenched and
sign-quenched reweighting.  The average reweighting factors in these
two schemes are given by the average phase factor and the average sign
of the fermion determinant in the phase-quenched and sign-quenched
ensembles, respectively,
\begin{align}
  R^{\det}_\text{pq}
  &=\ev{e^{iN_f\theta}}_{|{\det}^{N_f} D|} \,,\qquad
  R^\text{det}_\text{sq}
  =\langle \sign(\re{\det}^{N_f}D)\rangle_{|\re\,{\det}^{N_f} D|} \,,
  \label{Rdsq}
\end{align}
where we used $\det D=|\det D|e^{i\theta}$.  Note that
$R^{\det}_\text{pq}$ is the generalization of \eqref{phasepq} to
arbitrary $N_f$.  The reweighting factors $R^\text{det}_\text{pq}$ and
$R^\text{det}_\text{sq}$ can efficiently be computed numerically with
the subset method as explained in ref.~\cite[section~IV.C]{Bloch:2012ye}.

\begin{figure}
\centering
\includegraphics{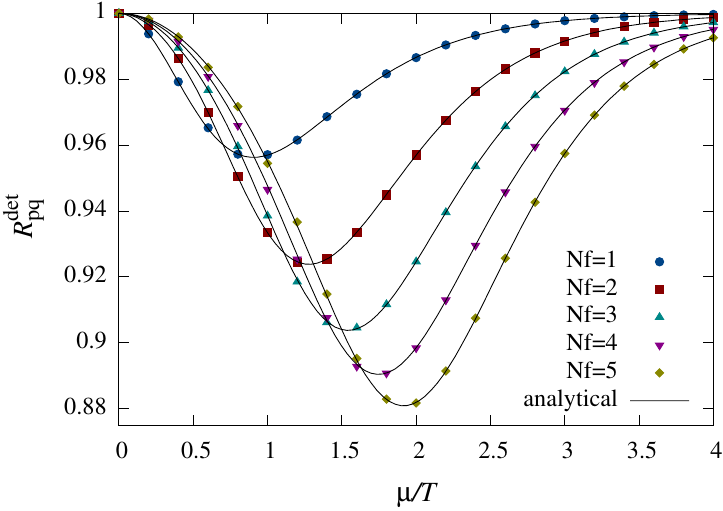}\hfill
\includegraphics{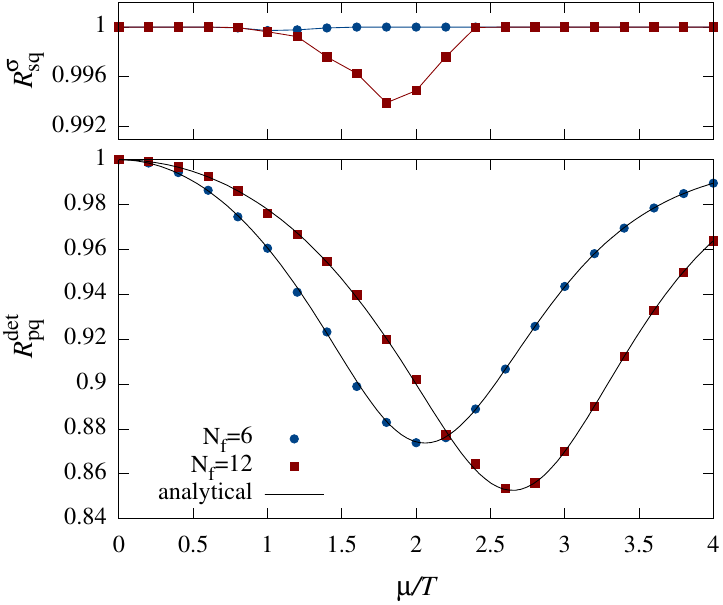}
\caption{Average phase factor
  $R^\text{det}_\text{pq}=\ev{e^{iN_f\theta}}_{|\det^{N_f}D|}$ of the
  fermion determinant in the phase-quenched ensemble (at vanishing
  mass) for $N_f=1,2,3,4,5$ (left) and $N_f=6,12$ (bottom right). The
  symbols represent the data measured with the subset method, the
  solid lines are the analytical results from
  appendix~\ref{app:avgphase}.  The narrow window at the top of the
  right panel shows the average sign
  $R^\sigma_\text{sq}=\ev{\sign\sigma}_{|\sigma|}$ of the subset
  weights, illustrating that there is a mild sign problem for the
  subset method when $N_f\geq6$. }
\label{fig_factors}
\end{figure}

Let us first discuss phase-quenched reweighting in the link
formulation.  We derived analytical formulas for the average phase
$R^{\det}_\text{pq}$ in the phase-quenched ensemble, which can be
expressed as the ratio of the unquenched and phase-quenched partition
functions, see eq.~\eqref{eq:pqNf}. For even $N_f$ the phase-quenched
partition function can easily be computed, as it can be expressed as
an integral over quarks and conjugate quarks. For odd $N_f$ the
computation is less trivial, as it involves an absolute
value. Nevertheless, we were able to calculate analytical expressions
for any $N_f$ in the massless case, see appendix~\ref{app:avgphase}.
The numerical and analytical results for $R^{\det}_\text{pq}$ are
shown in figure~\ref{fig_factors}. The numerical results were computed
using the subset method. For $N_f=1$ to $5$ (left panel) the subset
method can be used as is, but for $N_f=6$ and $N_f=12$ (right panel) a
mild sign problem develops, and we use sign-quenched reweighting on
the subset method, see \eqref{eq:sqrsm}. In this case we also show the
average sign $R_\text{sq}^\sigma$ of the subsets in the small top
window. The sign problem in the subset method is clearly much milder
than in the phase-quenched reweighting scheme using the SU(3) links,
as the average sign is much closer to one.  Note that the minima in
the bottom and top windows are shifted, meaning that the subset method
has no sign problem where the sign problem of the phase-quenched
reweighting scheme is maximal.

\begin{figure}\centering
\includegraphics{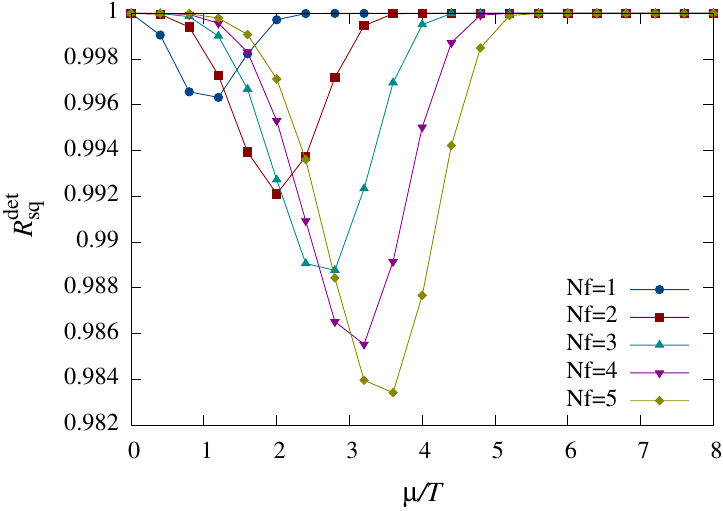}\hfill
\includegraphics{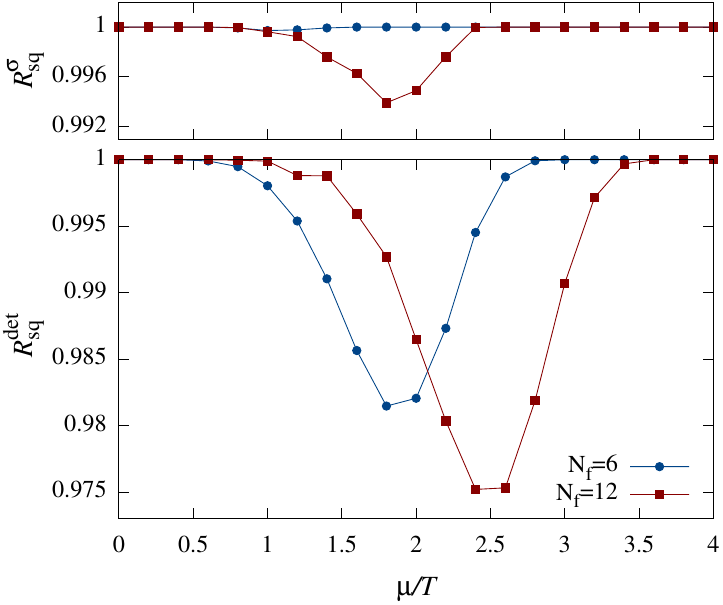}
\caption{Average sign
  $R^\text{det}_\text{sq}=\ev{\sign(\re\det^{N_f}D)}_{|\re\det^{N_f}D)|}$
  of the real part of the fermion determinant in the sign-quenched
  ensemble (at vanishing mass) for $N_f=1,2,3,4,5$ (left) and
  $N_f=6,12$ (bottom right). Again, the narrow top window in the right
  panel shows the average sign of the subsets.  The lines are drawn to
  guide the eye.}
\label{fig_factors_two}
\end{figure}

Let us now turn to sign-quenched reweighting in the link formulation,
where the reweighting factor is the average sign of the real part of
the determinant, rather than that of the subset weights, see
eq.~\eqref{Rdsq}. While analytical results are not available for this
average sign, numerical results can easily be obtained with the subset
method and are shown in figure~\ref{fig_factors_two}. Even though the
average sign of the determinant is closer to one than its average
phase \cite{deForcrand:2002pa}, the subset reweighting scheme is still
clearly superior.

\begin{figure}
\centerline{\includegraphics{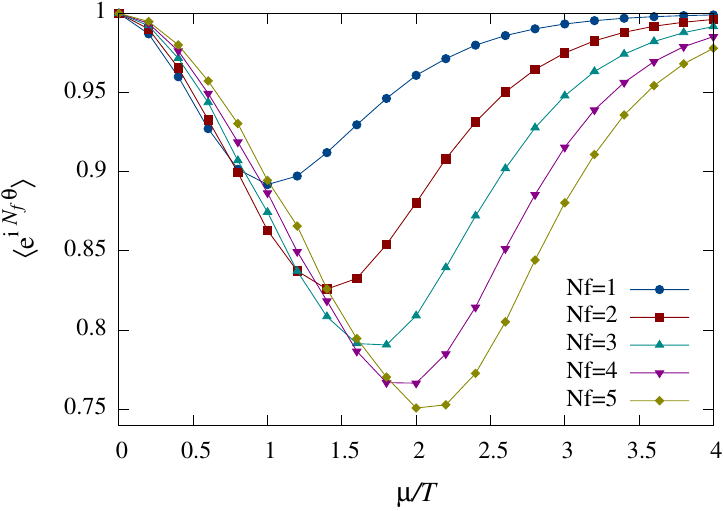}}
\caption{Average phase $\langle e^{iN_f\theta}\rangle$ of the fermion
  determinant in the unquenched ensemble as a function of $\mu/T$
  for $N_f=1,2,3,4,5$ with $m=0$.}
\label{Fig:phase}
\end{figure}

Besides computing the average phase of the fermion determinant in the
phase-quenched theory, which is relevant in the analysis of the
phase-quenched reweighting scheme, one can also investigate this
average phase in the full dynamical ensemble. An analytical expression
for this quantity is currently not available, but numerical results
can readily be obtained with the subset method.  We show these results
in figure~\ref{Fig:phase}. As expected, the average phase is smaller
in the full theory than in the phase-quenched theory.

\subsection{Extended subsets}

Although the sign problem in the subset method is still mild for $N_f
\geq 6$ and its severity only grows slowly with increasing $N_f$ it
would be preferable to avoid negative subset weights altogether. To
achieve this we now extend the subset construction beyond the mere
$Z_3$ rotations introduced in section~\ref{sec:subsets}.

Different routes can be taken. One could take advantage of the
invariance of the Haar measure under rotation by a constant group
element, which implies
\begin{align}
\int dP \, f(P) = \int dP \, f(G P) 
\label{Haarinv}
\end{align}
for arbitrary $G \in \SU(3)$. The \qone partition function can then be
rewritten as an integral over extended subsets $\Omega^G_{P} =
\Omega_P \cup \Omega_{GP}$, where $\Omega_P$ and $\Omega_{GP}$ are
constructed following eq.~\eqref{eq:subscc}.\footnote{This construction
  can be generalized to more than one group element, i.e.,
  $\Omega^{\{G\}}_{P} = \Omega_P \cup \Omega_{G_1P} \cup \Omega_{G_2P}
  \cup\ldots$} The sign problem would then be solved if a constant
group element $G$ could be found for which the sum of determinants of
the configurations in $\Omega^G_{P}$ is nonnegative.  However, it is
not clear how to find such a $G$ as it requires an analysis of the
landscape of the subset weight $\sigma(\Omega_P)$ in the full
8-dimensional parameter space of $P$.

What seems like a more feasible task is to analyze the landscape of
the subset weight in the eigenvalue representation
$P=\diag(e^{i\theta_1},e^{i\theta_2},e^{-i\theta_1-i\theta_2})$, see
appendix~\ref{app:eigrep}, and try to find a constant $G$ of the form
$G=\diag(e^{i\alpha},e^{i\beta},e^{-i\alpha-i\beta})$ to solve the
sign problem.  To see why this is sensible, first note that any $P$
can be diagonalized as
\begin{align}
  P = U\diag(e^{i\theta_1},e^{i\theta_2},e^{-i\theta_1-i\theta_2})\,U^\dagger\,.
\end{align}
We now shift the angles,
\begin{align}
  \theta_1\to\theta_1'=\theta_1+\alpha\,,\qquad
  \theta_2\to\theta_2'=\theta_2+\beta\,,
\end{align}
and create a ``rotated'' link
\begin{align}
  P' = R(P,G) \equiv U\diag(e^{i\theta_1'},e^{i\theta_2'},
  e^{-i\theta_1'-i\theta_2'})\,U^\dagger
  \label{rotP}
\end{align}
that has the same eigenvectors as $P$.  It is then straightforward to
show that the partition function can be rewritten as
\begin{align}
  Z^{(N_f)} &= \frac12 \left[ \int dP\, \sigma(\Omega_P) +
    \int dP'\, \sigma(\Omega_{P'}) \right] \notag\\
  &= \frac12 \int dP \left[ \sigma(\Omega_P) +
    \frac{J(\theta_1',\theta_2')}{J(\theta_1,\theta_2)} \,
    \sigma(\Omega_{P'}) \right] ,
  \label{ZPP'}
\end{align}
where $J$ is given in \eqref{eq:J}.  The contribution of $P'$ is
rescaled by the ratio of the Jacobians of $P'$ and $P$.  Now note that
in \qone the subset weight is a class function, i.e., it only depends
on $\theta_1$ and $\theta_2$, rather than the full $P$.  We can
therefore rewrite \eqref{ZPP'} as
\begin{align}
  Z^{(N_f)} &= \frac12 \int J(\theta_1,\theta_2)\,d\theta_1\theta_2 
  \left[ \sigma(\Omega_P) +
    \frac{J(\theta_1',\theta_2')}{J(\theta_1,\theta_2)} \,
    \sigma(\Omega_{P'}) \right] .
  \label{ZPG}
\end{align}
For observables that are also class functions, it suffices to generate
diagonal Polyakov loops according to the weight
$J(\theta_1,\theta_2)$.  We numerically verified the validity of both
\eqref{ZPP'} and \eqref{ZPG} by computing observables in both
formulations.

The extension of the original $Z_3$ subset by an additional $Z_3$
subset constructed with a single $G$ does not obey the symmetry of
$\tr P$ under permutations of the angles $\theta_1$, $\theta_2$, and
$\theta_3=-\theta_1-\theta_2$. Therefore we consider a larger
extension of the subset using six rotations
$\mathcal{G}=\{G_1,\ldots,G_6\}$ with all possible permutations of
$\alpha$ and $\beta$ over the three eigenvalues, i.e.,
\begin{equation}
  \begin{split}    
    G_1&=\diag(e^{i\alpha},e^{i\beta},e^{-i\alpha-i\beta})\,,\quad
    G_2=\diag(e^{-i\alpha-i\beta},e^{i\alpha},e^{i\beta})\,,\quad
    G_3=\diag(e^{i\beta},e^{-i\alpha-i\beta},e^{i\alpha})\,,\\
    G_4&=\diag(e^{i\beta},e^{i\alpha},e^{-i\alpha-i\beta})\,,\quad
    G_5=\diag(e^{-i\alpha-i\beta},e^{i\beta},e^{i\alpha})\,,\quad
    G_6=\diag(e^{i\alpha},e^{-i\alpha-i\beta},e^{i\beta})\,.
  \end{split}
\end{equation}
If $\alpha=\beta$, $-\beta/2$, or $-2\beta$ the number of permutations
is reduced to three. The extended subsets, containing 21
configurations and their complex conjugates, are defined as
\begin{align}
  \Omega^\text{ext}_P = \bigcup_{i=0}^6\Omega_{P^{(i)}}
\end{align}
with $P^{(i)}= R(P,G_i)$ using the rotations \eqref{rotP}, and
$G_0=\one$. The \qone partition function can then be rewritten as
\begin{align}
  Z^{(N_f)}= \int dP \,\sigma^\text{ext}_{P}
\end{align}
with extended subset weights
\begin{align}
  \sigma^\text{ext}_{P} = \frac17 \sum_{i=0}^6 \frac{J(P^{(i)})}{J(P)}
  \, \sigma(\Omega_{P^{(i)}}) \,,
\label{extweights}
\end{align}
where $J(P)=J(\theta_1,\theta_2)$ and $\sigma(\Omega_{P^{(i)}})$ is
the $Z_3$ subset weight defined in eq.~\eqref{sigmaNf}.  The subset
measurement on the extended subset is defined as
\begin{align}
  \langle O \rangle_{\Omega^\text{ext}_{P}} 
  &= \frac1{7\sigma^\text{ext}_{P}} \sum_{i=0}^6 
  \frac{J(P^{(i)})}{J(P)}\, \sigma(\Omega_{P^{(i)}}) \, 
  \langle O \rangle_{\Omega_{P^{(i)}}}
\end{align}
with the measurement on a single $Z_3$ subset as in \eqref{eq:meas}.  With
this definition, \eqref{<O>} and \eqref{Obar} straightforwardly
generalize to extended subsets (and any $N_f$).

\begin{figure}
  \includegraphics[width=0.48\textwidth]{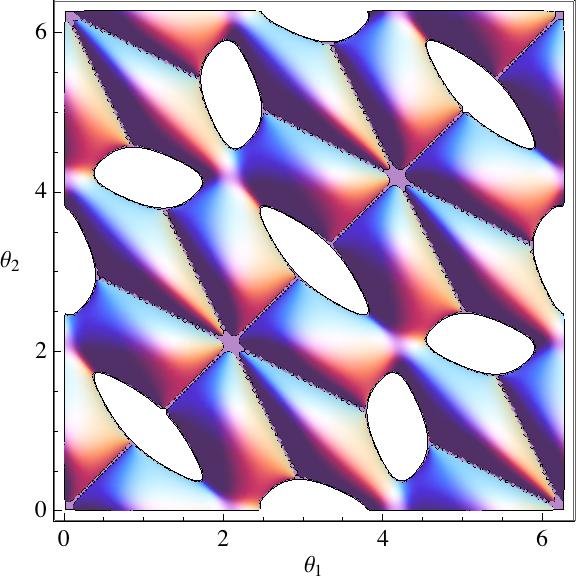}\hfill
  \includegraphics[width=0.48\textwidth]{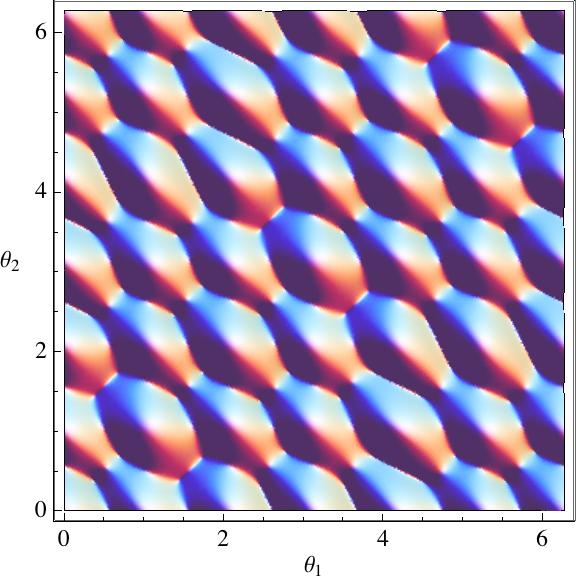}\\[1mm]
  \includegraphics[width=0.48\textwidth]{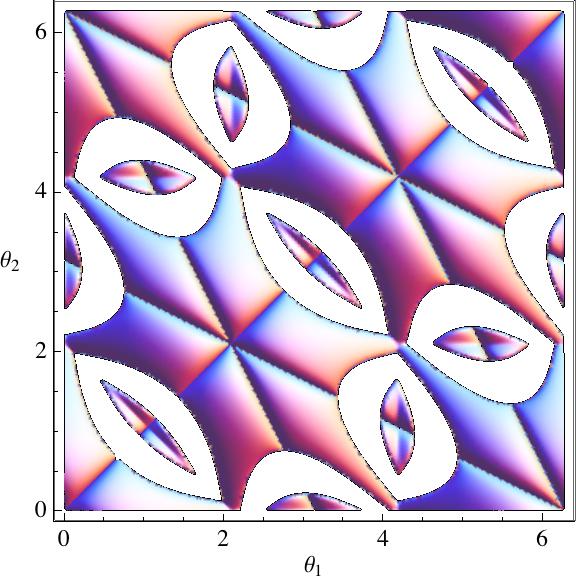}\hfill
  \includegraphics[width=0.48\textwidth]{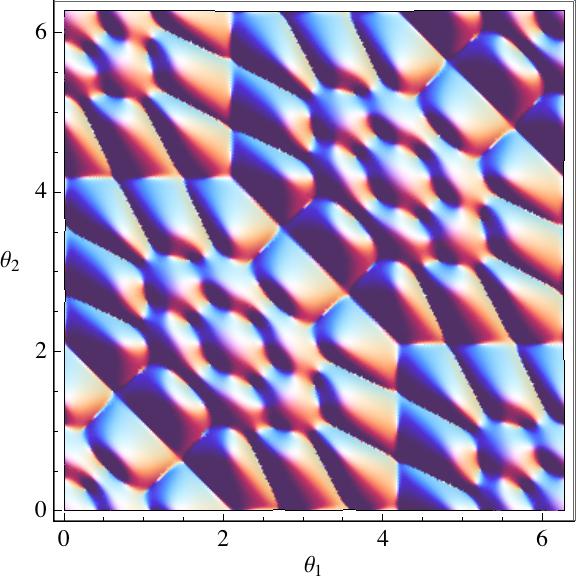}\\[-8mm]
  \caption{Plot of $\log [J(P)\,\sigma(\Omega_P)]$ (left) and $\log
    [J(P)\,\sigma^\text{ext}_P]$ (right) as a function of the Polyakov
    loop angles for $\mu/T=2.6$ (top) and $1.3$ (bottom), both for
    $N_f=24$ and $m=0$. The holes in the surface correspond to
    negative weights on the logarithmic scale. On the left, the
    weights \eqref{sigmaNf} of the original $Z_3$ subsets exhibit a
    clear sign problem. On the right, the weights \eqref{extweights}
    of the extended subsets were generated using $\alpha=-\beta=\pi/3$
    for $\mu/T=2.6$ and $\alpha=-\beta=\pi/4$ for $\mu/T=1.3$,
    respectively.  We observe that the sign problem is eliminated for
    these parameter values.}
  \label{fig:sigma_Nf=24}
\end{figure}

In figure~\ref{fig:sigma_Nf=24} we give an example of the effect of
the rotations $G_i$ on the extended subset weight for $N_f=24$ and
$\mu/T=2.6$ (top) and $\mu/T=1.3$ (bottom).\footnote{Note that
  allowing $\theta_1$ and $\theta_2$ to range from 0 to $2\pi$ leads
  to a mosaic of six replicated regions, caused by the permutation
  symmetries of the subset weights. The fundamental region can, for
  example, be defined as the triangle with corners
  $(\theta_1,\theta_2) \in \{(2\pi/3,2\pi/3),
  (4\pi/3,4\pi/3),(2\pi,0)\}$ delimited by the lines
  $\theta_2=\theta_1$, $2\theta_2+\theta_1=2\pi$, and
  $\theta_2+2\theta_1=4\pi$.}  Note that the sign problem is maximal
for $\mu/T=2.6$, see figure~\ref{fig:avgsgn} below.  In the left
panels we show the logarithm of $J(P)$ times the subset weights
\eqref{sigmaNf} for the original $Z_3$ subsets.  We clearly see
regions where the subset weights are negative.  The Swiss-cheese
pattern is fairly rigid in the sense that the location of the holes
seems independent of $N_f$ and $\mu$; only their sizes change, and
occasionally some isle-formation is observed inside the
holes.\footnote{The average sign $R_\text{sq}^\sigma$ equals $0.98836$
  for $\mu/T=2.6$ and $0.99988$ for $\mu/T=1.3$, i.e., the sign
  problem is much milder in the latter case although the holes are of
  similar size.  In this case the absolute value of the weights in the
  negative region is very small compared to the positive region.}
From the location of the holes we can make an educated guess for good
shifts $\alpha$ and $\beta$ for the construction of the rotations
$G_i$.  In the right panels of figure~\ref{fig:sigma_Nf=24} we show
the logarithm of $J(P)$ times the subset weights \eqref{extweights}
for extended subsets with $\alpha$ and $\beta$ given in the caption.
The extended subsets solve the sign problem in the illustrated cases.
For large $N_f$ there does not seem to be a single choice
$\mathcal{G}$ of rotations that solves the sign problem over the
complete $\mu$ range, and one should adapt $\mathcal{G}$ to the value
of $\mu$.\footnote{Even if we have picked a $\mathcal{G}$ that does
  not solve the sign problem completely the numerical simulations can
  simply switch to sign-quenched reweighting if negative weights
  occur.  The average sign can be used to monitor the remaining (very
  mild) sign problem.}

As a final illustration of the extended subsets we compare the average
sign of the original $Z_3$ subset weight to the average sign of the
extended subsets with a fixed choice of $\alpha=-\beta=\pi/3$ as a
function of $\mu/T$ for $N_f=12,24,48$ in figure~\ref{fig:avgsgn}.
The $Z_3$ subsets have a mild sign problem, which is completely
eliminated by the extended subsets for $N_f=12$ and eliminated for
almost all values of $\mu/T$ for $N_f=24$ and $48$ (for the fixed
choice of $\alpha$ and $\beta$).
\begin{figure}
  \centering
  \includegraphics{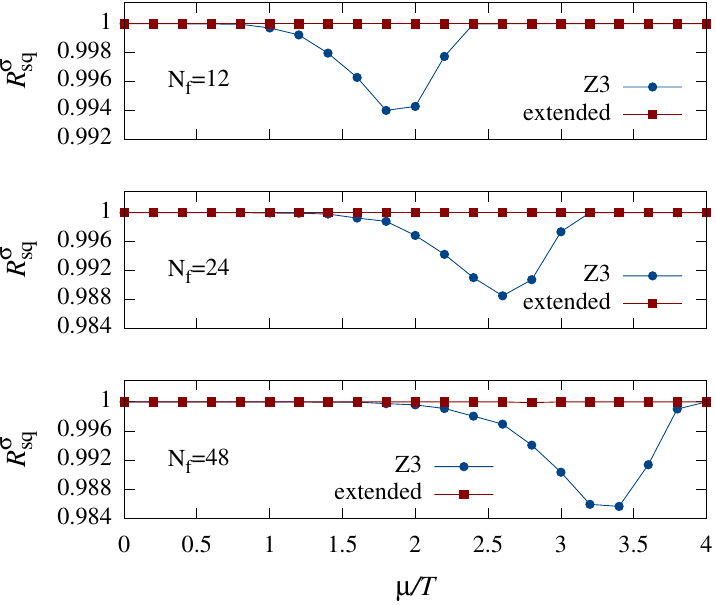}
  \caption{Comparison of the average sign $R^\sigma_\text{sq}$ of the
    subset weight for the original $Z_3$ subsets and for the extended
    subsets, with $N_f=12, 24, 48$, $m=0$, and fixed
    $\alpha=-\beta=\pi/3$.  The occasional tiny deviations from 1 for
    the extended subsets are hardly visible.}
  \label{fig:avgsgn}
\end{figure}

We should remark that our subset solution to the sign problem in \qone
does not immediately yield a solution to the exponential sign problem
occurring in higher dimensions ($d>1$), even if the resulting subset
weights are positive.  Indeed, if we assume naively that each temporal
link requires the construction of its own ``one-link subset'' with $N$
elements, we would have $N^V$ elements per ``lattice subset'' for a
$d$-dimensional lattice with volume $V=N_t\times N_s^{d-1}$.  In that
case the number of elements, and thus the computation time, grows
exponentially with the volume.  To solve the exponential sign problem,
a subset solution on a higher-dimensional lattice should have a
\emph{collective} nature, where subsets are formed using collective
rather than individual transformations of the links on different sites
such that the higher-dimensional subset is not a direct product of
subsets on each lattice site.

\section{Summary}
\label{sec:summ}

In this paper we presented a subset method to eliminate the sign
problem in dynamical simulations of QCD in 0+1 dimensions at nonzero
chemical potential. The SU(3) links are gathered into subsets, each
containing three links that are related by a $Z_3$ rotation. We showed
that the sum of fermion determinants of the three configurations in
any such subset is real and positive for $N_f=1$. For $N_f=2,\ldots,5$
the real parts of the subset weights are still positive.  Their
imaginary parts, which can be nonvanishing, are removed by adding the
complex conjugate links to the subsets such that the subsets contain
six links in total. The positive subset weights can then be used to
generate Markov chains of relevant subsets using importance
sampling. The method was illustrated by computing the quark number
density, the chiral condensate, and the Polyakov loop numerically and
comparing the data with analytical results, some of which were derived
for the first time in this paper.

For $N_f\geq 6$ the subset weights are no longer generically positive.
However, their average sign is still very close to unity so that
reweighting methods can be used.  We showed that sign-quenched
reweighting in the subset ensemble is much more efficient than the
standard phase-quenched or sign-quenched reweighting on the individual
determinants.

Finally, we constructed extended subsets by performing additional
SU(3) rotations. If chosen judiciously, these extended subsets have
positive weights even for $N_f \geq 6$.

Although we managed to get rid of the non-positivity of the weights in
the partition function by gathering configurations into subsets, we
cannot claim to have solved an exponential sign problem.  A creative
adaptation of the subset construction to higher dimensions is needed
to arrive at such a solution.

\acknowledgments{We thank Christof Gattringer and Kim Splittorff for
  useful discussions. This work was supported in part by the DFG
  collaborative research center SFB/TRR-55, by DFG grant BR 2872/4-2,
  and by the Alexander von Humboldt Foundation.}

\appendix

\section{One-flavor determinant}
\label{app:detD}

In this appendix we compute the coefficients $D_q$ in expression
\eqref{eq:decomp} for the $N_f=1$ determinant.  We start with
eq.~\eqref{detD} without the prefactor $1/2^{3N_t}$,
\begin{align}
  \det D = \det (e^\mu \, P + e^{-\mu} \, P^\dagger + A \, \one_3) \,,
\end{align}
where for simplicity of notation we replaced $\mu/T\to\mu$.  This
determinant can be computed explicitly using the formula for $3\times
3$ determinants in terms of traces, 
\begin{align} 
  6\det M= (\tr M)^3-3\tr M \tr M^2+2\tr M^3 \,,
\end{align}
where $M=e^\mu \, P + e^{-\mu} \, P^\dagger + A \, \one_3$ in this
case. Using the traces
\begin{align}
  \tr M &= e^\mu \, \tr P + 3 A + e^{-\mu} \, \tr P^\dagger  \,, \\
  \tr M^2 &= e^{2\mu} \, \tr P^2 + e^\mu \, 2 A\tr P + 3 A^2 + 6  + e^{-\mu} \, 2 A \tr P^\dagger + e^{-2\mu} \tr {P^\dagger}^2 \,,\\
  \tr M^3 &= e^{3\mu} \tr P^3 + e^{2\mu} \, 3 A \tr P^2 + e^\mu \, 3 (A^2+1) \tr P + 3 A^3 + 18A \notag\\
  &\quad + e^{-\mu} \, 3 (A^2+1) \tr P^\dagger + e^{-2\mu} \, 3 A \tr {P^\dagger}^2 + e^{-3\mu} \tr {P^\dagger}^3 
\end{align}
we find
\begin{align}
  \det D = \sum_{q=-3}^3 D_q \, e^{q\mu}
  \label{appdetD}
\end{align}
with coefficients
\begin{align}
D_0(P) &= A^3 - 3A + A |\tr P|^2 \,,\label{D0}\\
D_1(P) &= D_{-1}(P)^* = (A^2-2)\tr P + (\tr P^\dagger)^2 \,,\label{D1}\\
D_2(P) &= D_{-2}(P)^* = A\tr P^\dagger \,,\label{D2}\\
D_3(P) &= D_{-3}(P)^* = 1 \,,\label{D3}
\end{align}
where we also used $6\det P = (\tr P)^3-3\tr P \tr P^2+2\tr P^3 = 6$
as $P \in \SU(3)$, and $\tr P^2 = (\tr P)^2 - 2\tr P^\dagger$, which
follows, e.g., from \eqref{eq:eigrep}.

\section{Some integrals of traces}
\label{app:traces}

In our calculations we need some integrals of powers of traces of the
(anti-) Polyakov loop.  To compute them, we construct the tensor
product of $k$ copies of the fundamental and $\ell$ copies of the
anti-fundamental representation of SU(3) and decompose the product
into irreducible representations,
\begin{align}
  t=\bset{k\text{ times}}{3\otimes\cdots\otimes3}\otimes
  \bset{\ell\text{ times}}{\bar3\otimes\cdots\otimes\bar3}
  =\bigoplus_i n_ir_i\,,
\end{align}
where $t$ stands for the tensor product representation and $n_i$ is
the multiplicity with which the irreducible representation $r_i$
occurs in the decomposition.  We now take the trace of $P$ in the
tensor product representation and obtain
\begin{align}
  \label{eq:traces}
  \int dP\,\tr P^{(t)}=\int dP\,(\tr P)^k(\tr P^\dagger)^\ell
  =\sum_in_i\bset{=\delta_{i0}}{\int dP\,\tr P^{(r_i)}}=n_0\,,
\end{align}
where $r_0$ is the trivial representation and the last integral
follows from the orthonormality relations of the group characters.
Note that \eqref{eq:traces} is only nonzero if $(k-\ell)\bmod3=0$.  By
suitable choices of $k$ and $\ell$ we obtain
\begin{align}
  \int dP\, |\tr P|^2&=1 \quad\text{from}\quad
  3\otimes\bar3 = 8\oplus 1 \,,\\
  \int dP\, (\tr P)^3&=1 \quad\text{from}\quad
  3\otimes 3\otimes 3 = 10 \oplus 8 \oplus 8 \oplus 1 \,,\\
  \int dP\, |\tr P|^4&=2 \quad\text{from}\quad
  3\otimes\bar3\otimes3\otimes\bar3 = 27\oplus10\oplus\bar{10} \oplus
  8 \oplus 8 \oplus 8 \oplus 8\oplus 1 \oplus 1 \,.
\end{align}
Alternatively, we could use the eigenvalue representation of $P$ (see
appendix \ref{app:eigrep}) and integrate over $\theta_1$ and
$\theta_2$.  

\section{Relation between quark number density and Polyakov loop}
\label{app:nP}

In this appendix we consider a general number $N_f$ of flavors.  The
quark number density is
\begin{align}
  n &= \frac{T}{N_f}\frac{\partial\log Z^{(N_f)}}{\partial\mu} =
  \frac{1}{Z^{(N_f)}}\int dP\, {\det}^{N_f}D\, \tr\gamma\,,
\end{align}
where $\gamma$ is defined as
\begin{align}    
  \gamma &=
  \frac{p-1/p}{p+1/p+2\cosh(\muct)}
  \label{nNf}
\end{align}
with $p=e^{\mut} P$.  Here and below we have replaced
$\mu_c/T\to\mu_c$ and $\mu/T\to\mu$ for simplicity of notation.  We
also assume $\mu_c\ge0$ without loss of generality.  After factorizing
the denominator of $\gamma$ and performing a partial fraction
expansion we find
\begin{align}
\gamma
&=\frac{e^{\pm\muct}(p-1/p)}{(1+e^{\pm\muct}p)(1+e^{\pm\muct}/p)} \notag\\
&= \frac{1}{1+e^{\mupmm} P^\dagger} - \frac{1}{1+e^{\mupmp} P} \,.
\label{gamma}
\end{align}
Although it is not obvious from the final expression, \eqref{gamma} is
valid for either sign in front of $\mu_c$.  However, the same sign has
to be chosen for all occurrences of $\mu_c$.  Each of the two terms in
\eqref{gamma} can be considered as the result of a geometric
series. When expanding these series care has to be taken to ensure
their convergence. As the eigenvalues of $P$ all have unit magnitude,
the convergence requires the exponents in \eqref{gamma} to be
negative. The signs of these exponents depend on the relative values
of $\mu$ and $\mu_c$, for which we distinguish three cases that are
detailed in table~\ref{Table:geom}.
\begin{table}
\centerline{
\begin{tabular}{|c|c|c|c|}
\hline
& (a) & (b) & (c) \\
& $\mu < -\mu_c$ & $-\mu_c<\mu<\mu_c$ & $\mu_c<\mu$\\
\hline
$+\mu_c-\mu$ & + & + & -- \\
$+\mu_c+\mu$ & -- & + & + \\
\hline
$-\mu_c-\mu$ & + & -- & -- \\
$-\mu_c+\mu$ & -- & -- & + \\
\hline
\end{tabular}
}
\caption{Signs of the exponents $\pm\mu_c\pm\mu$ in \eqref{gamma} for
  $\mu<-\mu_c$, $-\mu_c<\mu<\mu_c$, and $\mu_c<\mu$.} 
\label{Table:geom}
\end{table}
From the table we read off that the following expansion is valid for
$-\mu_c<\mu<\mu_c$:
\begin{align}
\gamma
&= \sum_{\omega=1}^\infty (-)^\omega \left( e^{\omega\mumm}(P^\dagger)^\omega - e^{\omega\mump}P^\omega \right).
\label{(b)}
\end{align}
To expand \eqref{gamma} in a convergent series for $\mu<-\mu_c$ we
first extract $e^{\mupmm} P^\dagger$ from the denominator of
the first term and find
\begin{align}
\gamma &= \frac{e^{\mumpp} P}{1+e^{\mumpp} P} - \frac{1}{1+e^{\mupmp} P} 
= - 1 - 2 \sum_{\omega=1}^\infty  (-)^\omega  \cosh(\omega\muct) \, e^{\omega\mut} P^\omega .
\label{(a)}
\end{align}
Analogously, the series for $\mu_c<\mu$ is given by
\begin{align}
\gamma &= \frac{1}{1+e^{\mupmm} P^\dagger} - \frac{e^{\mumpm} P^\dagger}{1+e^{\mumpm} P^\dagger} 
= 1 + 2 \sum_{\omega=1}^\infty (-)^\omega \cosh(\omega\muct) \, e^{-\omega\mut} (P^\dagger)^\omega .
\label{(c)}
\end{align}
From the eigenvalue representation of the Polyakov loop we can show
(by counting powers of $e^{i\theta_1}$ and $e^{i\theta_2}$) that the
expectation value of $\tr P^{\pm\omega}$ is zero for $|\omega|>2N_f+3$
such that substitution of eqs.~\eqref{(b)}--\eqref{(c)} in
eq.~\eqref{nNf} yields
\begin{align}
n &=  
\begin{cases}
-3 - 2 \sum_{\omega=1}^{2N_f+3}  (-)^\omega  \cosh(\omega\muct) \, e^{\omega\mut} \langle\tr P^\omega\rangle\,, & \mu<-\mu_c\,,\\[1mm]
 \sum_{\omega=1}^{2N_f+3} (-)^\omega e^{-\omega\muct} \left( e^{-\omega\mut}\langle\tr(P^\dagger)^\omega\rangle - e^{\omega\mut}\langle\tr P^\omega\rangle \right), & -\mu_c < \mu < \mu_c\,, \\[1mm]
3+ 2 \sum_{\omega=1}^{2N_f+3} (-)^\omega \cosh(\omega\muct) \, e^{-\omega\mut} \langle\tr (P^\dagger)^\omega \rangle\,, &\mu_c < \mu \,.
\end{cases}
\label{finsums}
\end{align}
We conjecture that all three finite sums are identical over the
complete $\mu$-range, which we explicitly verified for $N_f=1$ and
$N_f=2$. If this is the case, a somewhat more symmetric equation is
found by combining the first and third formula,
\begin{align}
n &=  
 \sum_{\omega=1}^{2N_f+3}  (-)^\omega  \cosh(\omega\muct) \, 
 \left( e^{-\omega\mut} \langle\tr (P^\dagger)^\omega \rangle -
   e^{\omega\mut} \langle\tr P^\omega\rangle \right) .
\end{align}
Note that even though the finite sums \eqref{finsums} stand out by the
simplicity of their coefficients (for arbitrary $N_f$), shorter sums
involving lower winding numbers $\omega$, but with typically more
complicated coefficients, also exist. For instance, for $N_f=1$ a
simpler result is given in \eqref{eq:nP1}, while for $N_f=2$ we found
\begin{align}
n_{N_f=2} &= \frac{3}{2\!+\!\cosh2\muct} 
\Big[ \cosh\muct \langle e^{\mut}\tr P-e^{-\mut}\tr P^\dagger\rangle 
+ \frac14\langle e^{2\mut}\tr P^2-e^{-2\mut}\tr (P^\dagger)^2\rangle \Big] .
\end{align}
These short sums reduce the maximal winding number for $N_f=1$ from 5
to 1 and for $N_f=2$ from 7 to 2. Similar sums can be derived for
larger $N_f$, but no simple formula was found for their coefficients.

\section{SU(3) eigenvalue representation}
\label{app:eigrep}

It is often convenient to work in the eigenvalue representation of the
Polyakov loop.  The latter is diagonalized as $P=U\Lambda U^\dagger$
with
\begin{align}
  \label{eq:eigrep}
  \Lambda = \diag(e^{i\theta_1},e^{i\theta_2},e^{-i\theta_1-i\theta_2})\,,
\end{align}
where $\theta_1,\theta_2\in[0,2\pi]$ and $U\in\U(3)/\U(1)^{\otimes3}$.
The Haar measure is then given by
\begin{align}
  \label{eq:redHaar}
  dP&=J(\theta_1,\theta_2)\,d\theta_1\,d\theta_2\,dU\,,
\end{align}
where $dU$ is the normalized measure of $\U(3)/\U(1)^{\otimes3}$ and
the Jacobian is a Vandermonde determinant that only depends on
$\theta_1$ and $\theta_2$.  It is given by
\begin{align}
  J(\theta_1,\theta_2)
  &=\frac1{24\pi^2}\big|(e^{i\theta_1}-e^{i\theta_2})
  (e^{i\theta_1}-e^{-i\theta_1-i\theta_2})
  (e^{i\theta_2}-e^{-i\theta_1-i\theta_2})\big|^2 \notag\\
  &= \frac8{3\pi^2}\sin^2\frac{\theta_1-\theta_2}{2}
  \sin^2\frac{2\theta_1+\theta_2}{2}\sin^2\frac{\theta_1+2\theta_2}{2}
  \,, \label{eq:J}
\end{align}
where the prefactor ensures $\int dP=1$.

\section{Average phase in the phase-quenched theory}
\label{app:avgphase}

In this section we compute the average phase of the ${N_f}$-flavor
determinant in the phase-quenched theory.  This average phase factor
can be written as the ratio of the unquenched and phase-quenched
partition functions,
\begin{align}
  \label{eq:pqNf}
  \ev{e^{iN_f\theta}}_\text{pq}
  =\frac{\ev{(\det D)^{N_f}}_{N_f=0}}
  {\ev{|\det D|^{N_f}}_{N_f=0}} \,.
\end{align}
For odd $N_f$ it is difficult to compute the denominator on the RHS
since it involves square roots.  In the eigenvalue representation of
$P$ the fermion determinant is given by
\begin{align}
  \det D=\prod_{k=1}^3 (A+e^{\mu+i\theta_k}+e^{-\mu-i\theta_k})
  \quad\text{with}\quad \theta_3=-\theta_1-\theta_2\,,
\end{align}
where we again replaced $\mu/T\to\mu$ for simplicity of notation.
For a single angle we have
\begin{align}
  \big|A+e^{\mu+i\theta}+e^{-\mu-i\theta}\big|
  &=\big|A+2\cosh\mu\cos\theta+2i\sinh\mu\sin\theta\big|\notag\\
  &=\left[(2\cosh\mu+A\cos\theta)^2+(A^2-4)\sin^2\theta\right]^{1/2}.
\end{align}
This expression will be raised to the power $N_f$ so that the square
root disappears for even $N_f$.  For odd $N_f$ the square root cannot
be taken unless $A=2$, which corresponds to the chiral limit.  For
simplicity we now consider only this limit, in which
\begin{align}
  \big|A+e^{\mu+i\theta}+e^{-\mu-i\theta}\big|
  =2(\cosh\mu+\cos\theta)\,.
\end{align}
We can then compute $\evz{|\det D|^{N_f}}$ by integrating over
$\theta_1$ and $\theta_2$,
  \begin{align}
    \evz{|\det D|^{N_f}}&=8^{N_f}\iint_0^{2\pi}
    J(\theta_1,\theta_2)\,d\theta_1d\,\theta_2 \notag\\
    &\qquad\quad \times[(\cosh\mu+\cos\theta_1)(\cosh\mu+\cos\theta_2)
    (\cosh\mu+\cos(\theta_1+\theta_2))]^{N_f}.
\end{align}
In the following we state some explicit results for $\evz{(\det
  D)^{N_f}}$ and $\evz{|\det D|^{N_f}}$ in the chiral limit that have
been used in figure~\ref{fig_factors}.  For even $N_f$ these results
can easily be generalized to $A>2$, while the possibility of such a
generalization is not obvious for odd $N_f$.
\begin{align*}
  \evz{\det D}&=4+2\cosh(3\mu)\,,\\
  \evz{|\det D|}&=2+2\cosh\mu+2\cosh(3\mu)\,,\\
  \evz{(\det D)^2}&=50+40\cosh(3\mu)+2\cosh(6\mu)\,,\\
  \evz{|\det D|^2}&=22+24\cosh\mu+20\cosh(2\mu)+16\cosh(3\mu)
  +8\cosh(4\mu)+2\cosh(6\mu)\,,\\
  \evz{(\det D)^3}&=980+980\cosh(3\mu)+112\cosh(6\mu)+2\cosh(9\mu)\,,\\
  \evz{|\det D|^3}&=330+612\cosh\mu+432\cosh(2\mu)+370\cosh(3\mu)
  +180\cosh(4\mu)\notag\\
  &\quad+90\cosh(5\mu)+40\cosh(6\mu)+18\cosh(7\mu)+2\cosh(9\mu)\,,\\
  \evz{(\det D)^4}&=24696+28224\cosh(3\mu)+5040\cosh(6\mu)
  +240\cosh(9\mu)+2\cosh(12\mu)\,,\\
  \evz{|\det D|^4}&=8434+15040\cosh\mu+12904\cosh(2\mu)
  +9360\cosh(3\mu)+6262\cosh(4\mu)\notag\\
  &\quad+3264\cosh(5\mu)+1832\cosh(6\mu)+720\cosh(7\mu)+272\cosh(8\mu)\notag\\
  &\quad+80\cosh(9\mu)+32\cosh(10\mu)+2\cosh(12\mu)\,,\\
  \evz{(\det D)^5}&=731808+914760\cosh(3\mu)+217800\cosh(6\mu)
  +18150\cosh(9\mu)\notag\\
  &\quad+440\cosh(12\mu)+2\cosh(15\mu)\,,\\
  \evz{|\det D|^5}&=241752+464250\cosh\mu
  +390980\cosh(2\mu)+308390\cosh(3\mu)\notag\\
  &\quad+209600\cosh(4\mu)+132706\cosh(5\mu)+72800\cosh(6\mu)\notag\\
  &\quad+37450\cosh(7\mu)+15540\cosh(8\mu)+6550\cosh(9\mu)
  +2100\cosh(10\mu)\notag\\
  &\quad+650\cosh(11\mu)+140\cosh(12\mu)+50\cosh(13\mu)+2\cosh(15\mu)\,,\\
  \evz{(\det D)^6}&=24293412+32391216\cosh(3\mu)+9447438\cosh(6\mu)
  +1145144\cosh(9\mu)\notag\\
  &\quad+52052\cosh(12\mu)+728\cosh(15\mu)+2\cosh(18\mu)\,,\\
  \evz{|\det D|^6}&=8100348+15385104\cosh\mu+13537836\cosh(2\mu)
  +10757544\cosh(3\mu)\notag\\
  &\quad+7905636\cosh(4\mu)+5210856\cosh(5\mu)+3192856\cosh(6\mu)\notag\\
  &\quad+1735920\cosh(7\mu)+875016\cosh(8\mu)+386960\cosh(9\mu)\notag\\
  &\quad+160974\cosh(10\mu)+55440\cosh(11\mu)+18832\cosh(12\mu)
  +5040\cosh(13\mu)\notag\\
  &\quad+1332\cosh(14\mu)+224\cosh(15\mu)+72\cosh(16\mu)+2\cosh(18\mu)\,,\\
  \evz{(\det D)^{12}}&=114801908084920000
  +183683052935872000\cosh(3\mu)\notag\\
  &\quad+93809559177963200\cosh(6\mu)
  +30350151498752800\cosh(9\mu)\notag\\
  &\quad+6136979163350750\cosh(12\mu)
  +759997419610000\cosh(15\mu)\notag\\
  &\quad+55999809866000\cosh(18\mu)
  +2357886731200\cosh(21\mu)\notag\\
  &\quad+53588334800\cosh(24\mu)
  +605176000\cosh(27\mu)\notag\\
  &\quad+2990000\cosh(30\mu)
  +5200\cosh(33\mu)
  +2\cosh(36\mu)\,,\\
  \evz{|\det D|^{12}}&=38228935544196544
  +74588093808767136\cosh\mu\notag\\
  &\quad+69256334029071024\cosh(2\mu)
  +61191395371363712\cosh(3\mu)\notag\\
  &\quad+51446943705501036\cosh(4\mu)
  +41144000777152416\cosh(5\mu)\notag\\
  &\quad+31295396191130192\cosh(6\mu)
  +22628417756712768\cosh(7\mu)\notag\\
  &\quad+15550003072387944\cosh(8\mu)
  +10148555203557056\cosh(9\mu)\notag\\
  &\quad+6288313704156072\cosh(10\mu)
  +3695893600669920\cosh(11\mu)\notag\\
  &\quad+2059587221609088\cosh(12\mu)
  +1086906242675088\cosh(13\mu)\notag\\
  &\quad+542935067374728\cosh(14\mu)
  +256303460885472\cosh(15\mu)\notag\\
  &\quad+114289127660268\cosh(16\mu)
  +48034918945728\cosh(17\mu)\notag\\
  &\quad+19022357453088\cosh(18\mu)
  +7076078360208\cosh(19\mu)\notag\\
  &\quad+2472827265612\cosh(20\mu)
  +808058178928\cosh(21\mu)\notag\\
  &\quad+247249295568\cosh(22\mu)
  +70294375008\cosh(23\mu)\notag\\
  &\quad+18651768124\cosh(24\mu)
  +4554250272\cosh(25\mu)
  +1036301112\cosh(26\mu)\notag\\
  &\quad+213188976\cosh(27\mu)
  +41112918\cosh(28\mu)
  +6918912\cosh(29\mu)\notag\\
  &\quad+1120264\cosh(30\mu)
  +144144\cosh(31\mu)
  +20880\cosh(32\mu)\notag\\
  &\quad+1456\cosh(33\mu)
  +288\cosh(34\mu)
  +2\cosh(36\mu)\,.
\end{align*}

\sloppypar
\bibliography{1dQCD}
\bibliographystyle{JBJHEP}

\end{document}